\documentclass[article,aps,pra,twocolumn,showpacs,superscriptaddress,groupedaddress]{revtex4}  
\usepackage{graphicx}  
\usepackage{dcolumn}   
\usepackage{bm}        
\usepackage{amssymb}   
\usepackage{amsmath}
\usepackage{amsmath}
\usepackage{braket}
\usepackage[caption=false]{subfig}
\usepackage{verbatim}
\usepackage{hyperref}
\usepackage{float}
\usepackage[none]{hyphenat}
\usepackage{eqnarray,amsmath}
\DeclareMathOperator{\Tr}{Tr}
\DeclareMathOperator{\Log}{log}
\usepackage{color}
\newcommand{\be}{\begin{equation}}
\newcommand{\ee}{\end{equation}}
\newcommand{\ben}{\begin{eqnarray}}
\newcommand{\een}{\end{eqnarray}}
\newcommand{\bes}{\begin{subequations}}
\newcommand{\ees}{\end{subequations}}
\newcommand{\bF}{\begin{figure}}
\newcommand{\eF}{\end{figure}}

\newcommand{\RNum}[1]{\uppercase\expandafter{\romannumeral #1\relax}}

\newcommand{\mbf}[1]{\mathbf{#1}}

\begin{document}

\title{Quantum tomography with random diagonal unitary maps and statistical bounds on information generation using random matrix theory}

\author{Sreeram PG}
\affiliation{Department of Physics, Indian Institute of Technology Madras, Chennai, India 600036}

\author{Vaibhav Madhok}
\affiliation{Department of Physics, Indian Institute of Technology Madras, Chennai, India 600036}

\begin{abstract}
We study quantum tomography from a continuous measurement record obtained by measuring expectation values of a set of Hermitian operators obtained from unitary evolution of an initial observable. For this purpose, we consider the application of a random unitary, diagonal in a fixed basis at each time step and quantify the information gain in tomography using Fisher information of the measurement record and the Shannon entropy associated with the eigenvalues of covariance matrix of the estimation.
 Surprisingly, very high fidelity of reconstruction is obtained using random unitaries diagonal in a fixed basis even though the measurement record is not informationally complete. We then compare this with the information generated and fidelities obtained by application of a different Haar random unitary at each time step.
 We give an upper bound on the maximal information that can be 
obtained in tomography and show that a covariance matrix taken from the Wishart-Laguerre ensemble of random matrices and the associated Marchenko-Pastur distribution saturates this bound. We find that physically, this corresponds to an application of a different Haar random unitary at each time step. We show that repeated application of random diagonal unitaries gives a covariance matrix in tomographic estimation that corresponds to a new ensemble of random matrices. 
We analytically and numerically estimate eigenvalues of this ensemble and show the information gain to be bounded from below by the Porter-Thomas distribution. 
 
\end{abstract}
\maketitle
\section{Introduction}
\begin{figure}[H]
\includegraphics[scale=0.4]{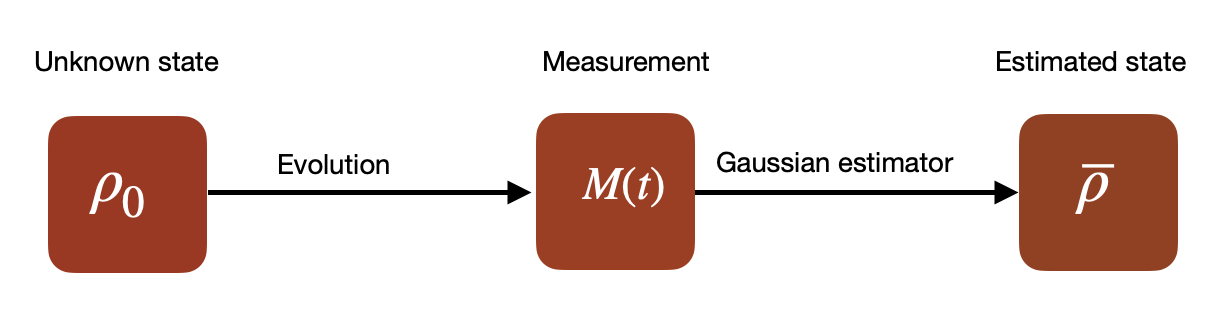}  \label{tom}
\caption{An overview of the state estimation procedure}
\end{figure}

To determine an unknown state is a  fundamental challenge in quantum information processing.
 The process of estimating an unknown state by performing measurements on it is called quantum tomography. Since the probabilities for various outcomes during the measurements depend on the state in which we perform them, we can in principle determine the unknown density  matrix by inverting the measurement records.  \cite{RevModPhys.29.74,paris2004quantum, dariano2003quantum}. 
 The traditional way to perform quantum tomography is to make projective measurements. 
   Any projective measurement would collapse the wave function, deterministically evolved through the Schr\"{o}dinger's equation. Projective measurements are expensive and time-consuming and one has to repeat the process many times to get an accurate estimate of the density matrix. 
 However, as an alternative, one may overcome this by employing the protocol
for tomography via weak continuous measurements \cite{PhysRevLett.95.030402, PhysRevLett.97.180403,chaudhury2009quantum,PhysRevLett.93.163602, PhysRevA.81.032126,riofrio2011continuous,Deutsch2010QuantumCA,PhysRevLett.124.110503,smith2003faraday,PhysRevA.90.032113}.
In this approach, the ensemble is collectively controlled and coherently evolved in a time-dependent manner to obtain an ``informationally complete" continuous measurement record. A set of measurement operators is called informationally complete, if they span all of the operator space.  Such a set of complete measurements has been extensively studied  \cite{renes2004symmetric,zhu2018universally,flammia2005minimal,scott2006tight,d2004informationally},  to cite a few. One does a series of measurements of several observables and obtains the outcome probabilities. Then one inverts these measurement records to obtain the original state.  An outline of the whole procedure  is shown in fig. \ref{tom}.
At a more fundamental level, continuous measurements provide us with a window to study quantum-to-classical transition, the  emergence of chaos from quantum mechanics, and information gain in tomography
under chaotic dynamics \cite{habib2006emergence,madhok2014information, bhattacharya2003continuous,  MADHOK2016}. 

  In this work, we study the connection between information gain in tomography and the randomness of quantum dynamics employed to generate the measurement record. For this purpose, we consider various families of ``random" maps.
  In particular, we consider the application of random unitaries diagonal in a fixed basis and compare this with applying a different Haar random map at each iteration. In the process, we obtain bounds on maximal information that can be acquired in for  state reconstruction. 
  
  Such diagonal unitaries are of natural interest in quantum computation. Experimentally diagonal gates can be fault-tolerantly realized in, e.g., super- and semi-conducting systems \cite{aliferis2009fault}. Diagonal quantum circuits are experimentally much simpler to implement and less sensitive to environmental decoherence  than non-diagonal quantum circuits { \cite{buscemi2007quantum}}.{ Further, the repeated action of diagonal unitaries has been shown to achieve decoupling of two interacting quantum systems \cite{nakata2017decoupling}. This could be used in achieving environmental decoherence.}  Since all gates in a diagonal circuit commute, 
  it enables us to realize the circuit by a single time-independent commuting
  Hamiltonian. Therefore, in an experimental realization of such dynamics, one does not need to worry about the order of interactions and it reduces the time for implementation and makes the protocol more robust.
  It has been shown that diagonal gates have better computational power than classical computers \cite{PhysRevLett.112.140505, bremner2010classical}. The entangling power of such unitaries has been also of interest \cite{Arul}. Diagonal quantum circuits have been employed in the generation of random quantum states uniformly distributed according to a unitarily invariant Haar measure 
  \cite{ nakata2014generating, nakata2014diagonal}.
  Despite this, the merits of
  random diagonal unitaries are far from exhausted and little is known about their concrete applications in other quantum information processing protocols like state reconstruction and quantum control.
  
  Our findings show that random unitary maps with diagonal unitaries do not lead to information completeness as far as the task of quantum state reconstruction is concerned. However, if the state to be reconstructed lies in a lower-dimensional subspace \cite{gross2010quantum} or if we only require a lower resolution tomography that serves practical purposes \cite{aaronson2007learnability}, implementing diagonal unitaries could not only be sufficient but also efficient. In this work, we do not put any restriction on the initial state or the resolution of tomography and study the information generation in state reconstruction when the underlying dynamics are random diagonal unitaries.
  
  Our study has an intimate connection with quantum chaos since the nature of chaotic dynamics can be effectively modeled by random maps. Probing this question deeper, we further explore whether the origin of information gain lies in the spectral statistics of the quantum chaotic map or in the randomness of its eigenvectors. To address this question, we study information gain obtained when the dynamics is generated by a map whose eigenvectors are random/chaotic but spectral statistics 
  belong to a Poisonnian distribution that is characteristic of a regular system. We also study the amount of information gain when eigenvalues are chosen from a distribution characteristic of a chaotic system.

  The remainder of the paper is organized as follows.  
In the next section, we describe a general protocol for tomography with continuous-time measurements. In Section \RNum{3} we use this protocol to study state reconstruction using random diagonal unitaries. We find that despite being a restrictive case, very good fidelities can be achieved. Then we quantify the information gain using Fisher information and Shannon entropy of eigenvalues of the covariance matrix of the estimation process in section \RNum{4}. We give predictions of information generation from random matrix theory, obtain bounds on information gain and introduce a new random matrix ensemble and show that our predictions agree with our numerical simulations in section \RNum{5}. 
We discuss the connection of information gain in continuous measurement tomography and quantum chaos to spectral statistics and eigenvectors of dynamical maps  \RNum{6}
before we conclude with a brief discussion and overview of our results in the final section.
  
%
  
  \section{Weak Continuous Measurements}
  The total system we work with is composed of the object system $(S)$ and the probe/meter $(M)$. We assume that the object system and the probe start out in a product state. Their evolution is governed by the total Hamiltonian $H_\tau =H_S+ H_M +H_{int}$ \cite{svensson2013pedagogical}. Here $H_{int}$ is the interaction Hamiltonian between the system and the meter. 
  The system and the meter are initially assumed to be uncoupled.  
  They evolve together via a time evolution operator ${U},$ generated by the total Hamiltonian, $H_\tau$.
  \begin{equation}
  {U} \tau_0 {U}^\dagger= {U} \sigma_0 \otimes \mu_0 {U}^\dagger
  \end{equation}
  where ${U}= \mathrm{exp} \left(- \frac{i}{\hbar} \int \mathrm{d}t H_\tau \right)$, $\sigma_0$ belong to the object system and $\mu_0$ to the probe.

  Let the initial state of the meter be $|m^{(0)}\rangle.$ After undergoing unitary evolution for time $t$, the state becomes $|m^{(i)}\rangle$. Expanding in terms of a continuous pointer variable $\hat{Q},$ with pointer states $|q \rangle$,
  \begin{equation}
  |m^{(0)}\rangle= \int \mathrm{d}q |q \rangle \langle q|m^{(0)}\rangle= \int \mathrm{d}q|q\rangle \psi_0(q)
  \end{equation}
  \begin{equation}
  |m^{(i)}\rangle= \int \mathrm{d}q |q \rangle \langle q|m^{(i)}\rangle= \int \mathrm{d}q|q\rangle \psi_i(q)
  \end{equation}
  Let us assume that initially, the pointer of the meter is centered around $q=0$ so that $\langle Q\rangle_0=0$. That is a natural and convenient choice because the difference in the pointer variable is what characterizes a measurement. Let us also choose the initial wave function of the meter to be a Gaussian, centered at zero. 
  \begin{equation}
  \psi_0(q)= \dfrac{1}{(2 \pi \sigma^2)^{1/4}} \mathrm{exp}\left(\frac{-q^2}{4 \sigma^2} \right)
  \end{equation}
  where $\sigma$ is the width of the Gaussian probability density. In the collective, weak measurement that we do, the collective observable say $\mathcal{O}_c= \sum \mathcal{O}^j$, where $\mathcal{O}^j$ acts on the $j^{th}$ subsystem. 
  The interaction Hamiltonian  $H_{int} = \gamma \mathcal{O}_c \otimes \hat{P}$ captures the coupling of the observable to be measured, with a  meter observable. The variable $\hat{P}$ is chosen to be the one conjugate to the pointer variable $\hat{ Q}.$ Here $\gamma$ is a  coupling constant. The measurement is supposed to occur during a short time interval $\delta t_u$, so that
  \begin{equation}
  \int \mathrm{d}t H_{int} = \int \mathrm{d}t \gamma \mathcal{O}_c \otimes \hat{P}= \gamma \mathcal{O}_c \otimes \hat{P} \delta t_u
  \end{equation}
  The combination $\gamma \delta t_u = g$ is an effective coupling constant.  In our case, we are driving the system using random unitaries. However,   the measurement procedure is only concerned with the interaction term in the Hamiltonian. Since our aim is is to explain the measurement process, for simplicity let us set $H_S$ and $H_M$ to zero. 
  Then
  \begin{equation}
  {U} = 
  \mathrm{exp} \left( \frac{-i}{\hbar}g \mathcal{O}_c \otimes \hat{P} \right)
  \end{equation}
  To understand how the coupled evolution changes the meter variable, assume that the system starts in a pure state  $\ket{\phi}_s=\sum_i \alpha_i|o_i\rangle$, where $\lbrace o_i \rbrace$ are the eigenstates of $\mathcal{O}_c$. Then the collective state after interaction is given by
  \begin{widetext}
  \begin{align}
  \sum_i \alpha_i |o_{i}\rangle \otimes |m^{(i)}\rangle &= {U} \left( \ket{\phi}_s \otimes |m^{(0)}\rangle \right)  \\&= 
  \mathrm{exp} \left( \frac{-i}{\hbar}g \mathcal{O}_c \otimes \hat{P} \right) \left(\Big(\sum_i \alpha_i|o_{i}\rangle \Big) \otimes |m^{(0)}\rangle \right) \\&= \sum_i \alpha_i|o_i\rangle \otimes \mathrm{exp} \left( \frac{-i}{\hbar}g o_i  \hat{P} \right) |m^{(0)}\rangle \\&= \sum_i \alpha_i |o_i\rangle \otimes \mathrm{exp} \left( \frac{-i}{\hbar}g o_i  \hat{P} \right) \int \mathrm{d}q |q\rangle \psi_0(q) \\&= \sum_i \alpha_i|o_i\rangle \otimes \int \mathrm{d}q |q\rangle \psi_0(q- go_i) \label{state}
  \end{align}
  \end{widetext}
  The meter state after evolution is, $|m^{(i)}\rangle = \int \mathrm{d}q |q\rangle \psi_0(q- go_i)$ with probability $|\alpha_i|^2$, which implies that $\psi_i(q) = \psi_0(q-go_i)$.
  That is, the initial pointer state of the meter has been translated proportional to  an eigenvalue of the system observable. 
  Until now a measurement of the meter hasn't been performed. Now let us perform a projective   measurement of the meter.
  If the meter is projected onto a particular outcome, then the rest is an operator acting on the system Hilbert space, called a Kraus operator \cite{nielsen2002quantum}.
  \begin{align}
  M_q&= \langle q| \mathrm{exp} \left( \frac{-i}{\hbar}g \mathcal{O}_c \otimes \hat{P} \right)|m_0\rangle\\&= \mathrm{exp} \left( \frac{-i}{\hbar}g \mathcal{O}_c \otimes \hat{P} \right) \psi_0(q) \\ &= \sum_i \psi_0(q-go_i) |o_i\rangle \langle o_i| \\&= \sum_i \dfrac{1}{(2 \pi \sigma^2)^{1/4}} \mathrm{exp}\left(\frac{-(q-go_i)^2}{4 \sigma^2} \right) |o_i\rangle \langle o_i| \label{measure}
  \end{align}
  Let $\rho_0$ denote the initial state of the object system. Then the  post measurement state is given by  
  \begin{equation}
  \rho_q'= \frac{M_q \rho_0 M_q^\dagger}{\mathrm{Prob}(q)}. \label{post}\end{equation} 
  The corresponding POVM element $ E_q = M_q ^{\dagger} M_q$ is given by
  \begin{equation}
  E_q =\sum_i \dfrac{1}{(2 \pi \sigma^2)^{1/2}} \mathrm{exp}\left(\frac{-(q-go_i)^2}{2 \sigma^2} \right) |o_i\rangle \langle o_i|
  \end{equation}
  Note that    $\mathrm{Lim}_{\sigma \rightarrow0}E_q= |go_i =q\rangle \langle go_i=q|$. For a finite $\sigma$ the measurement has finite strength. For large $\sigma$, the measurement is very weak.
  
  The probability for a measurement  outcome $q$ is given by $\mathrm{Prob}(q)= \mathrm{Tr}(E_q \rho_0).$  For the initial  state $\ket{\phi}_s=\sum_i \alpha_i|o_i\rangle$, we get
  \begin{equation}
  \mathrm{Prob}(q) =  \dfrac{1}{(2 \pi \sigma^2)^{1/2}}\sum_i |\alpha_i|^2 \mathrm{exp}\left(\frac{-(q-go_i)^2}{2 \sigma^2} \right) \label{prob}
  \end{equation}
  In the weak-measurement regime, when $\sigma \gg o_i$ holds for all eigenvalues,  probability function in \ref{prob}  can be rewritten as follows.
  \begin{align}
  \mathrm{Prob}(q) &=  \dfrac{1}{(2 \pi \sigma^2)^{1/2}}\sum_i |\alpha_i|^2 \mathrm{exp}\left(\frac{-(q-go_i)^2}{2 \sigma^2} \right) \\
  &\approx \dfrac{1}{(2 \pi \sigma^2)^{1/2}} \mathrm{exp}\left(\frac{-(q-\sum_i |\alpha_i|^2 go_i)^2}{2 \sigma^2} \right) \label{approx}\\
  &= \dfrac{1}{(2 \pi \sigma^2)^{1/2}} \mathrm{exp}\left(\frac{-(q- g \langle o_i\rangle)^2}{2 \sigma^2} \right) 
  \end{align}
  Equation \ref{approx} is obtained by Taylor expanding the exponential function up to first-order around $q=0$ \cite{vaidman1996weak}. The Gaussian spread,  $\sigma^2$ is called shot noise. There is also another noise arising due to the fundamental uncertainty in quantum measurements.  Fluctuations in the observed meter state, called the projection noise leads to variations in the system state.  However, since the shot noise $\sigma^2 $ is much larger, the effect of the projection noise can be neglected and   measurement induced back-action is insignificant \cite{smith2003faraday}.  Therefore it is possible to perform multiple measurements without needing to repeat the evolution from the start and  obtain a time-stamped series of measurement records.

  \section{Continuous measurement tomography}
     In the continuous measurement tomography protocol that we consider, one starts with an ensemble of $\mathcal{N}$, non-interacting, simultaneously prepared quantum systems in an identical but unknown state described by the density matrix $\rho_0^{\otimes \mathcal{N}}$, where $\rho_0$ is the density matrix of a single system. The ensemble is collectively controlled, coherently evolved, and continuously probed to obtain an ``informationally complete" continuous measurement record. In order to achieve information completeness,  the set of measured observables should span an operator basis for $\rho_0$ when viewed in the Heisenberg picture. For a Hilbert space of finite dimension $d$,  fixing the normalization of $\rho_0$, the set of Hermitian operators must form a basis of $su(d)$.
     
     We measure the sum of identical observables on all the $\mathcal{N}$ subsystems, and the measurement record at time $t$ can be written in terms of such a collective observable as
\begin{equation}
M(t)=  \langle \mathcal{O}_0 \rangle (t) + \delta M(t)
\end{equation}
Here, $\delta M(t) $ arises from the noise in the detection system, and $\langle \mathcal{O}_0 \rangle(t) = \mathrm{Tr}(\rho_0 U(t)^\dagger \mathcal{O}_0 U (t))$.
Our goal is to determine $\rho_0$ by continuously measuring an observable $\mathcal{O}_0$ evolved in the Heisenberg picture. Such a collective measurement in principle can lead to correlations \cite{geremia2005suppression} among the states which can cause back-action. However, under the conditions of weak continuous measurements, any such quantum back-action is negligible  \cite{PhysRevLett.95.030402}. The prominent noise in the system is the intrinsic shot noise of the probe.

 We consider a discrete set of measurements separated by the time interval $\Delta t$, of observables $\mathcal{O}_n= \Big(\prod_{i=1}^n U_i^\dagger (\Delta t)\Big) \mathcal{O}_0 \Big( \prod_{i=1}^nU_i(\Delta t)\Big)$, where a different unitary governs the evolution for each $\Delta t$ interval. The unitary evolution, which would produce an informationally complete set of observables is not unique.
 The question we ask is the following - How does the performance of tomography, as quantified by the fidelities obtained, depend on the nature of the unitary or the set of unitaries employed to evolve the system.
  For example, one can choose $U( \Delta t)$s from the set of Haar random unitaries \cite{mezzadri2006generate,ozols2009generate}, and apply them to get an informationally complete measurement record that is also unbiased over time. We shall refer to this kind of dynamics as the ``Haar random" case in this work.

 One can also obtain a sequence of measurement records from repeated application of the same fixed unitary chosen at random according to Haar measure \cite{easton}, i.e., $\prod_{i=1}^nU_i(\Delta t)=U_0^n(\Delta t)$. This way,  we obtain a one-parameter family of measurement records. Although not informationally complete, this produces high fidelity reconstruction \cite{PhysRevA.81.032126}. Repeated application of a single random unitary has been studied as a paradigm to explore quantum signatures of chaos \cite{ madhok2014information, MADHOK2016}. { Another way of driving the operator evolution is by choosing random unitaries that are all diagonal in a particular basis. There is an extra degree of freedom of phases in this case, unlike the previously described powers of a single unitary.}

Let us now discuss the estimation procedure briefly. Considering a stroboscopic time series of measurement records, at a time $ n \Delta t$, 
\begin{equation}
M_n=  \mathrm{Tr}(\mathcal{O}_n \rho_0) +\sigma W(n) \label{00}
\end{equation}
 where we treat the detector noise as Gaussian white noise $\delta M(t)= \sigma W(t)$. Here  $\sigma$ is the noise variance and $W(t)$ is a Wiener process {with mean zero and unit variance} \cite{durrett2019probability}.
 We can expand $\rho_0$ in a Hermitian basis consisting of  $(d^2-1)$ traceless operators $E_\alpha$ and the identity matrix \cite{MADHOK2016}.
 \begin{equation}
 \rho_0 =  \sum_{\alpha=1}^{d^2-1} r_\alpha E_\alpha + I/d \label{01}
 \end{equation}
 Using (\ref{01}) in (\ref{00}), 

\begin{align}
M_n &=  \sum_{\alpha=1}^{d^2-1} r_\alpha \mathrm{Tr}( \mathcal{O}_n E_\alpha ) + \sigma W(n)\\
& =  \sum_{\alpha=1}^{d^2-1} r_\alpha \tilde{\mathcal{O}}_{n\alpha} +\sigma W(n)
\end{align}
where $\mathrm{Tr}( \mathcal{O}_n E_\alpha )=\tilde{\mathcal{O}}_{n\alpha} $.
All such measurement records $\lbrace M_n \rbrace$ together can be written in a matrix form, 
 \begin{equation}
 \bold{\tilde{M}} =  \bold{\tilde{\mathcal{O}}} \bold{r} +\sigma \bold{W} \label{matrix}
 \end{equation}
 where $\bold{\tilde{M}}$ is a vector of measurement records.
Equation \ref{matrix} says that  the conditional probability of the measurement records given the underlying parameters is a Gaussian, 
 \begin{align}
 \bold{P}({\bold{\tilde{M}/r}}) &\propto \mathrm{exp}\left(-\frac{1}{2 \sigma^2}(\bold{\tilde{M}} - \bold{\tilde{\mathcal{O}}r})^T (\bold{\tilde{M}} - \bold{\tilde{\mathcal{O}}r})\right) \label{cond} \\
 & \propto \mathrm{exp}\left(-\frac{1}{2}(\bold{r}-r^{ML})^T  \bold{C^{-1}} (\bold{r}-r^{ML}) \right) \label{cov}
 \end{align}
Equation \ref{cov} can be obtained from (\ref{cond}), look at \cite{riofrio2011continuous} for a proof.  Here  the maximum likelihood estimate  vector $r^{ML}$ of the parameters $\lbrace\alpha\rbrace$ is the one which minimizes the exponent in the Gaussian \cite{PhysRevA.55.R1561}, given by
 \begin{equation}
 r^{ML} =  \sigma^2(\bold{\tilde{\mathcal{O}}}^T \bold{\tilde{\mathcal{O}}})^{-1} \bold{\tilde{\mathcal{O}}}^T \bold{\tilde{M}},
 \end{equation}
where the quantity {$  \sigma^2(\bold{\tilde{\mathcal{O}}}^T \bold{\tilde{\mathcal{O}}})^{-1}$ is called the covariance matrix, $\bold{C}$. Therefore, $r^{ML} = \mbf{C}\mbf{\tilde{\mathcal{O}}}^{T} \mbf{\tilde{M}}$.} The eigenvalues of $\mbf{C}^{-1}$ are  the signal-to-noise ratios with which we have measured different orthogonal directions in the operator space (given by its eigenvectors). 
 
In the absence of measurement noise, and when the inverse covariance matrix $\bold{C^{-1}}= \mbf{\tilde{\mathcal{O}}}^T \mbf{\tilde{\mathcal{O}}}/\sigma^2$ is full rank, the most likelihood estimate is given by $ \rho^{ML} = \sum\limits_{\alpha=1}^{d^2-1} r^{ML}_\alpha E_\alpha + I/d $. In presence of measurement noise, or when the measurement record is incomplete, $\rho^{ML}$ can have non-physical eigenvalues. Then one has to replace $\rho^{ML}$ by its closest physical density matrix, which can be obtained by minimizing the squared distance between the new estimate $\bar{r}$ and $r^{ML}$ \cite{tarantola2005inverse,MADHOK2016}.
\begin{equation}
\left \|r^{ML}  - \bar{r} \right\|^2 = (r^{ML} -\bar{r})^T (\bold{\tilde{\mathcal{O}}}^T \bold{\tilde{\mathcal{O}}}) (r^{ML} -\bar{r})
\end{equation}
 subject to the constraint $\sum\limits_{\alpha=1}^{d^2-1} \bar{r}_{\alpha } E_\alpha  + {I}/{d}  \geq 0$.
 \section{Continuous measurement tomography with random diagonal unitaries}
 We evolve the initial state using random unitaries diagonal in a fixed basis and generate a measurement record. In the Heisenberg picture, the operator evolves while the state remains the same.  After the first $\Delta t$ time interval, the  operator $\mathcal{O}_0$ changes to $U^{\dagger}(\Delta t) \mathcal{O}_0 U(\Delta t)$, where $U(\Delta t)=\sum_{j=1}^d e^{-i \phi_j} \ket{j} \bra{j} $. Since we will be indexing the unitaries as well, 
 we can rewrite this as $U_{m}(\Delta t)=\sum_{j=1}^d e^{-i \phi_{mj}} \ket{j} \bra{j} $, where $U_{m}(\Delta t)$ is the random diagonal unitary applied at time $m\Delta t$.
 
 Here the exponential phase factors $\phi_{mj}, \in [0, 2\pi]$, are chosen uniformly at random.  
 After $n$ time steps, 
 \begin{equation}
 \mathcal{O}_n = U_{n}^{\dagger}(\Delta t)U_{n-1}^{\dagger}...U_{1}^{\dagger}(\Delta t)\mathcal{O}_0U_{1}(\Delta t)U_{2}(\Delta t)...U_{n}(\Delta t) \label{002}
 \end{equation}
 which gives
 \begin{align}
 \mathcal{O}_n &= \sum_{j,k=1}^d e^{\sum_{m=1}^n {-i(\phi_{mj}-\phi_{mk})}} \bra{k}\mathcal{O}_0\ket{j} \ket{k}\bra{j} \\
&= \sum_{j,k=1}^d e^{-i(\Phi_{nj}-\Phi_{nk})} \bra{k}\mathcal{O}_0\ket{j} \ket{k}\bra{j}   \label{on}
 \end{align}
 where $\Phi_{nj}= \sum_{m=1}^n \phi_{mj}$ is the phase multiplying $j^{th}$ eigenvector after the evolution for time $n \Delta t$.
The operators $\left\lbrace \mathcal{O}_n\right\rbrace$ do not span all of the operator space.  Consider $G= \left\lbrace g \in \mathrm{su}(d) | U(t) g U^{\dagger}(t)=g\right\rbrace$, where $U(t)$ is a unitary diagonal in a particular basis considered, at time $t$.  Let $B= \left\lbrace g  \in G | \mathrm{Tr}(g \mathcal{O}_0)=0 \right\rbrace$. Then $\mathrm{Tr}(\mathcal{O}(t)g)=0, \forall g \in B$. Here $\mathcal{O}(t)$ represents the operator evolved by $U(t)$. $G$ is isomorphic to the Cartan subalgebra of $su(d)$, and the dimension of $G \geq d-1$. Therefore the dimension of the spanned space $\leq d^2-d+1$. This is very similar to arguments presented in \cite{PhysRevA.81.032126}, quantifying the dimension of the operator space spanned under repeated application of a single unitary map.  But can the random diagonal dynamics saturate this bound or do they span a strictly lower dimensional subspace? To see this, let us
 rewrite (\ref{on}) as follows
\begin{multline}
\mathcal{O}_n=\sum_{j=1}^{d} \bra{j}\mathcal{O}_0\ket{j} \ket{j}\bra{j} +\\ \sum_{j \neq k =1 }^d e^{-i(\Phi_{nj}-\Phi_{nk})} \bra{k}\mathcal{O}_0\ket{j} \ket{k}\bra{j} \label{03}
\end{multline}
and use the condition
\begin{equation}
\sum_{n=0}^{d^2-d} a_n \mathcal{O}_n =0 \: \mathrm{iff}\: a_n=0 \forall n \label{04}
\end{equation}
for linear independence of the the observables.
It has been shown that if the following  conditions are satisfied,  the set $\lbrace\mathcal{O}_n\rbrace$ is linearly independent \cite{PhysRevA.81.032126}.
\begin{enumerate}
\item $ \bra{i}\mathcal{O}_0\ket{j} \neq 0 \: \forall i,j$
\item $\Phi_{mj}-\Phi_{mk} \neq \Phi_{m'j'}-\Phi_{m'k'}, \forall (m,j,k) \neq (m',j',k')$\label{condition2}
\end{enumerate}
In random diagonal unitaries, we pick the eigenphases uniformly at random, therefore condition  \ref{condition2} is satisfied for any typical member.     That is, the set of observables generated using diagonal in a basis unitaries span a $d^2-d+1$ operator subspace almost always.
This makes intuitive sense. Kinematically speaking, repeated application of a single unitary should do as well as a set of diagonal random unitaries with a fixed basis. Our numerical simulations and the yield of tomography give further evidence of this.

\begin{figure}
\centering
\includegraphics[width=8cm,height=5.5cm,angle=0]{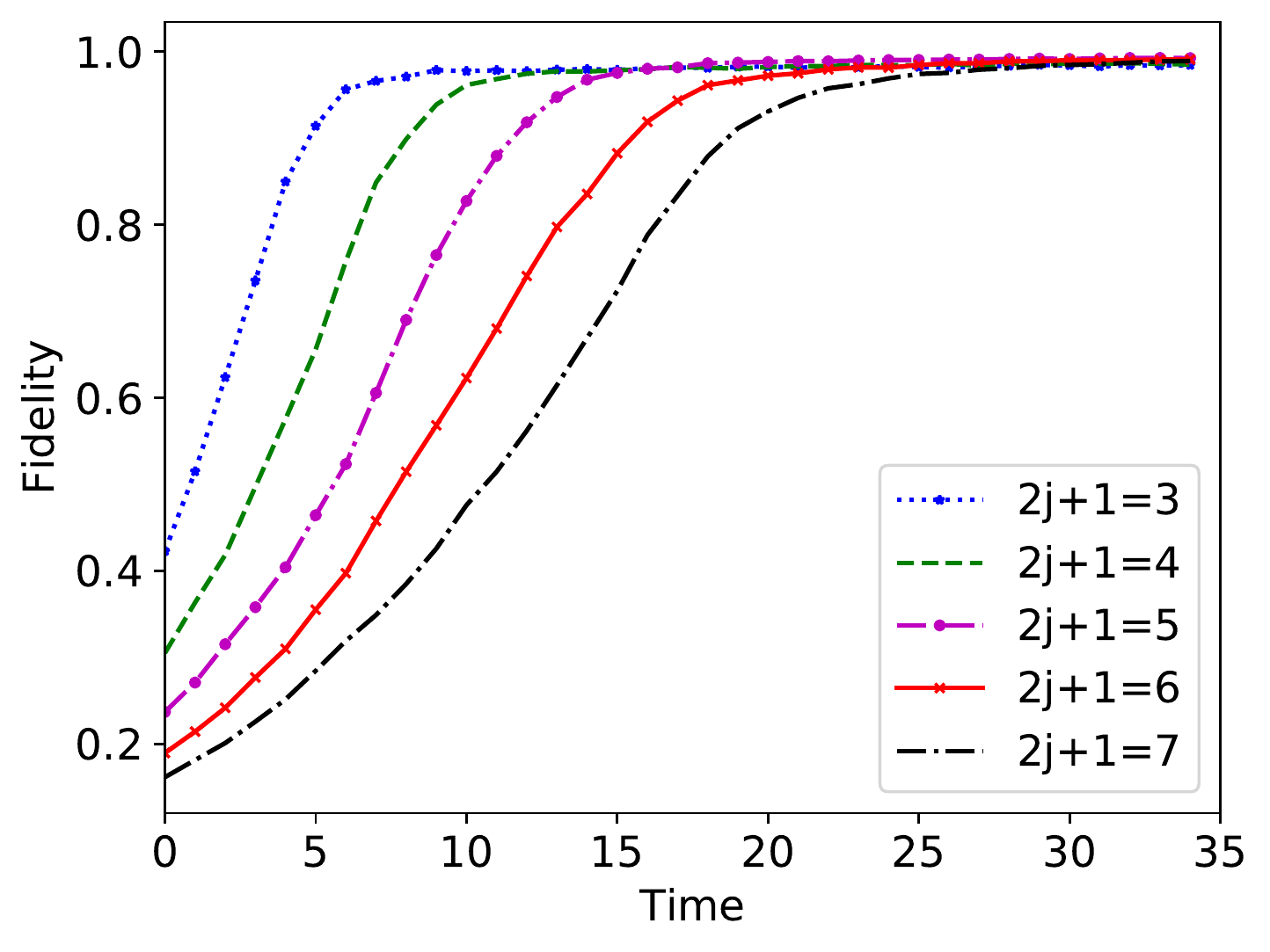} 
  \caption{Average fidelity of reconstruction with  random unitaries  diagonal in a fixed basis against time for different dimensions of Hilbert space. The X-axis represents number of applications of the unitary map. Averaging is done over reconstruction of 200 random pure states drawn according to Haar measure.  Figure shows that even for low dimensions, surprisingly high fidelity reconstruction $ >0.98$ is achieved even though measurement is not informationally complete. }
   \label{fig:1}
\end{figure}

\begin{figure*}
\subfloat[\label{sfig:testa}]{%
  \includegraphics[width=8cm,height=5.5cm,angle=0]{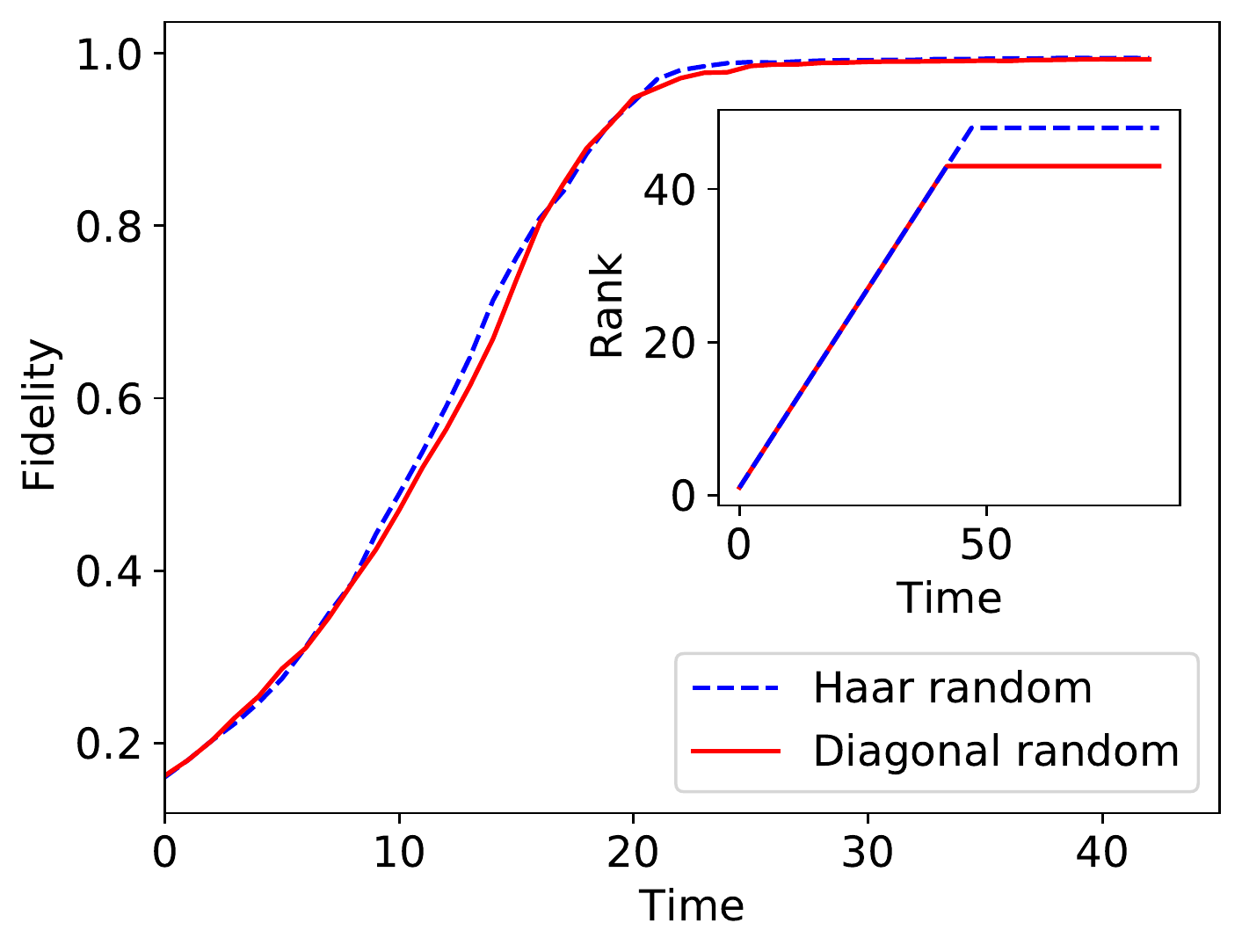}%
}\hfill
\subfloat[\label{sfig:testa}]{%
  \includegraphics[width=8cm,height=5.50cm,angle=0]{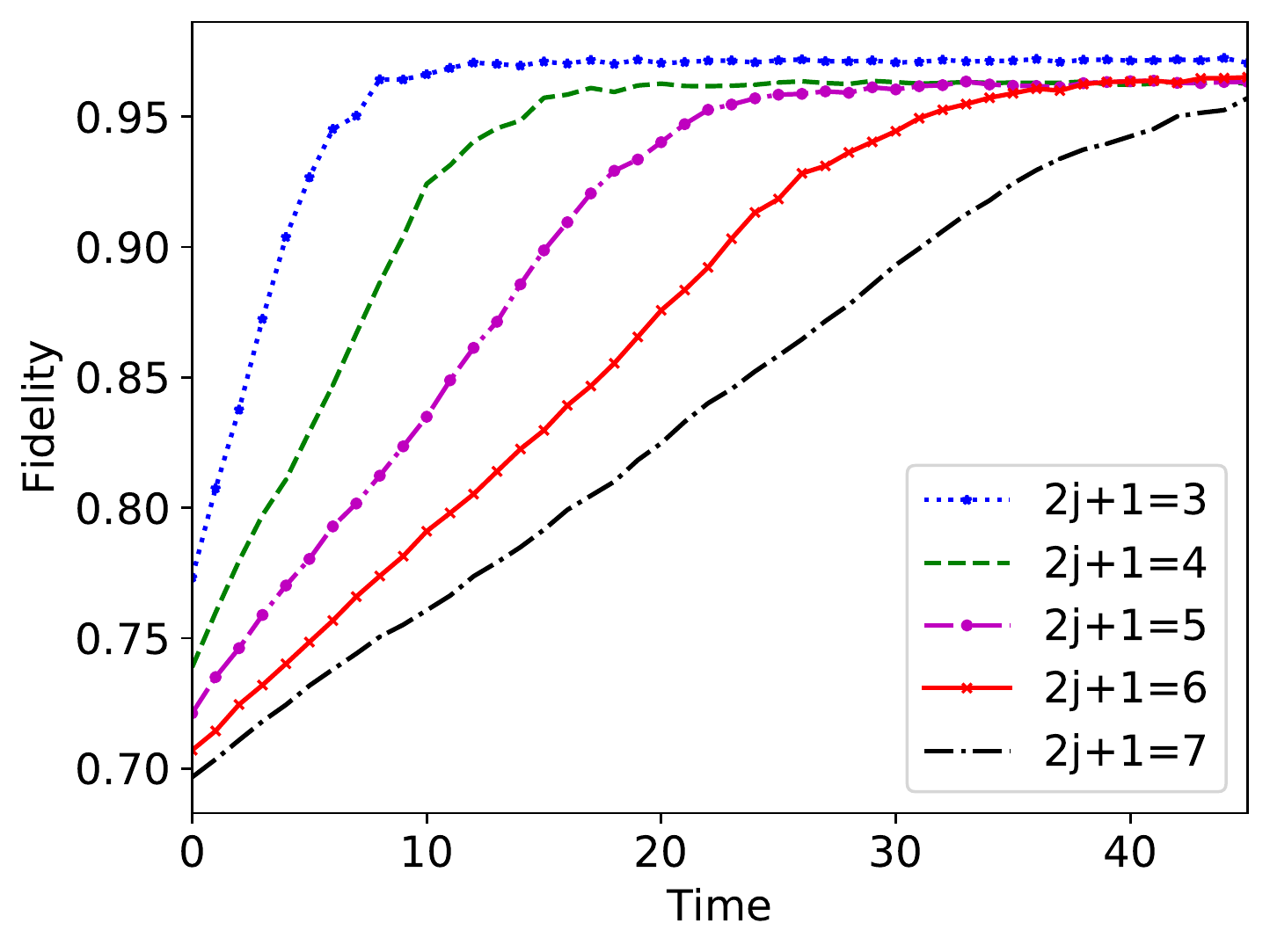}%
} 
\caption{a) Average fidelity of reconstruction  for $d=7$ with a random unitary  diagonal in a fixed basis and  completely Haar random unitary (i.e., a different Haar random unitary at each time step) against the number of applications of the unitary map. Averaging is done over the reconstruction of 100 random pure states drawn according to Haar measure. The rank of the covariance matrix against time is shown in the inset. It gives the dimension of the operator space spanned. Dynamics using   random unitary  diagonal in a fixed basis doesn't span all of the operator space, yet performance is similar. High fidelity is achieved even before rank saturates. b) Average fidelity of reconstruction against time, averaged over 200 mixed states picked according to Hilbert Schmidt measure for small dimensions. }

\label{fig:mix}
\end{figure*}

 Pure state performance when reconstructed with this algorithm for random unitary diagonal in a fixed basis evolution is shown in Fig. \ref{fig:1}. We see that with more measurement records, the reconstruction fidelity is increasing and saturating very close to one. As the Hilbert space dimension increases, the operator subspace about which we do not have any information becomes less significant and a near-complete reconstruction is achieved. However, it is remarkable that even for small dimensions, where one would expect the effect of the subspace not spanned to be more pronounced, fidelities $>$ 0.98 is achieved. But the same process for mixed states yields a noticeable difference in the reconstruction fidelities when random diagonals are used instead of Haar random unitaries, as seen in Fig. \ref{fig:mix}b.

\section{Quantifying Information gain in tomography}
To quantify the information gain during measurements,  consider the Hilbert Schmidt distance between the estimated state  $\tilde{\rho}$  and the actual state $ \rho$. 
\begin{equation}
e=
\Tr\left\lbrace(\tilde{\rho}-\rho)^2 \right\rbrace
\end{equation} 
It is easy to see that $e$ quantifies the error in the reconstruction \cite{PhysRevLett.88.130401}.
Using the expansion in (\ref{01}), the mean error $\langle e\rangle$, obtained by repeating the reconstruction procedure, can be expressed as
\begin{equation}
\left\langle e \right\rangle= \sum_{\alpha} \langle (\Delta r_\alpha)^2 \rangle \label{var}
\end{equation}
{where $\left\lbrace r_\alpha\right\rbrace$ are components of the state vector.}
The variances in (\ref{var}) are  bounded from below,  called the  Cramer Rao bound \cite{cramir1946mathematical} 
\begin{equation}
\langle (\Delta r_\alpha)^2 \rangle \geq \left[F^{-1}\right]_{\alpha\alpha} \label{fisher}
\end{equation}
where $F$ is the Fisher information matrix associated with the conditional distribution in (\ref{cond}).
 When there is negligible quantum back-action, all the uncertainty in a parameter value $r_\alpha$ is due to the shot noise variance $\sigma^2$,  and the Fisher information matrix equals the inverse covariance matrix  $F=\bold{C^{-1}}$\cite{vrehavcek2002invariant}.  
Now  looking at (\ref{fisher}),  the inverse of the total uncertainty can be written as  follows.\begin{equation} \frac{1}{\sum_\alpha \langle (\Delta r_\alpha)^2 \rangle}  = \frac{1}{\Tr(\bold{C})} \end{equation} $1/\Tr(\bold{C})$ can be intuitively understood as
 a measure of the net information gained from measurements,  called the collective Fisher information($FI$) \cite{MADHOK2016}.  It monotonically increases with more measurements as seen in Fig. \ref{entr}a.   Each eigenvector of $\bold{C^{-1}}$ represents an orthogonal direction in operator space that we have measured up to the final time, and each eigenvalue determines the information gain or  signal to noise ratio in that direction. {If the dynamics doesn't span all of the operator space, $\bold{C^{-1}}$ is not full rank, and  $FI$ is ill-defined. To rectify this situation, a Tikhonov regularization is performed by  adding a multiple of Identity to $\bold{C^{-1}}$ before inverting \cite{ng2004feature}.
\begin{figure*}
\subfloat[\label{sfig:testa}]{%
  \includegraphics[width=8cm,height=5.5cm,angle=0]{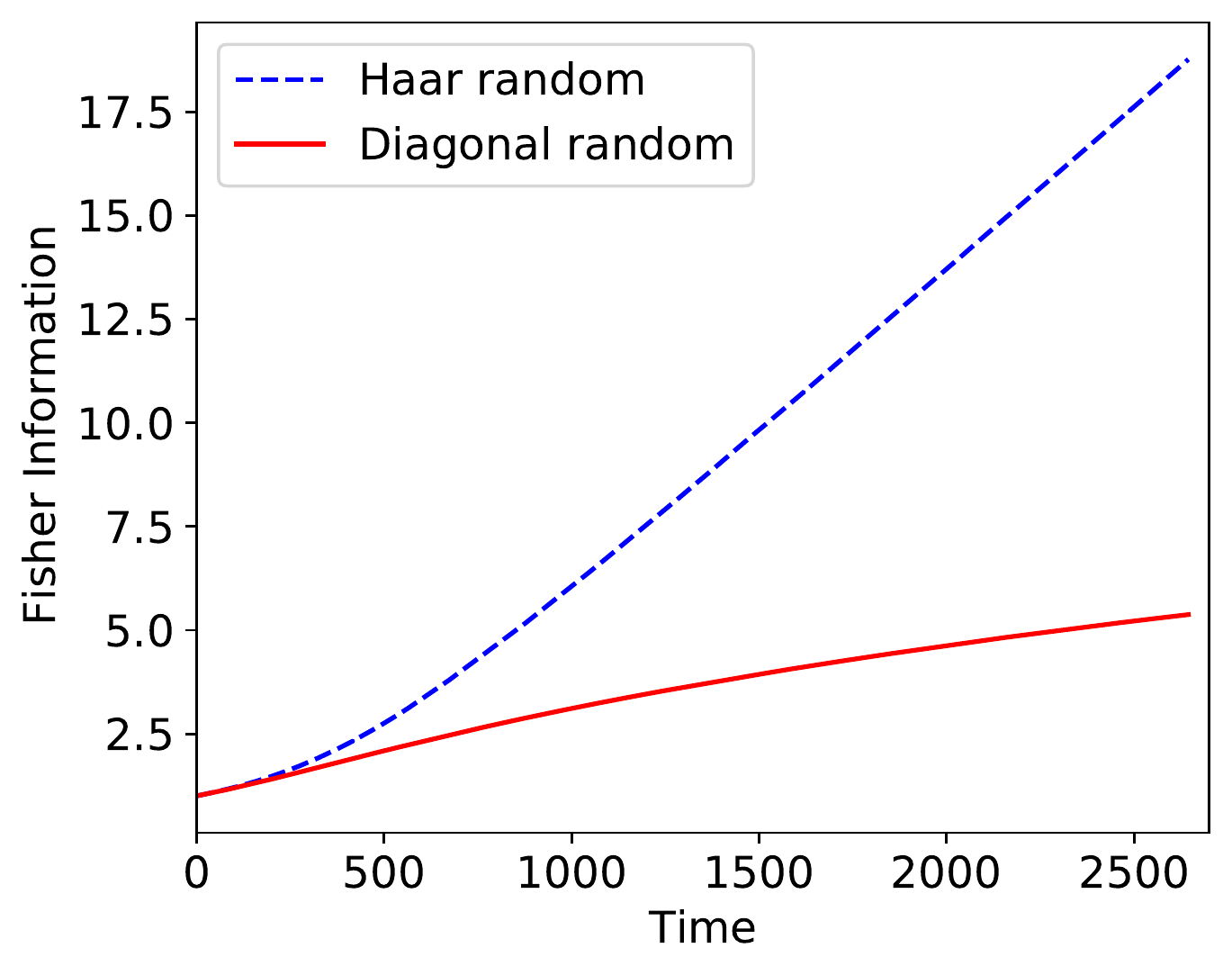}%
}\hfill
\subfloat[\label{sfig:testa}]{%
  \includegraphics[width=8cm,height=5.5cm,angle=0]{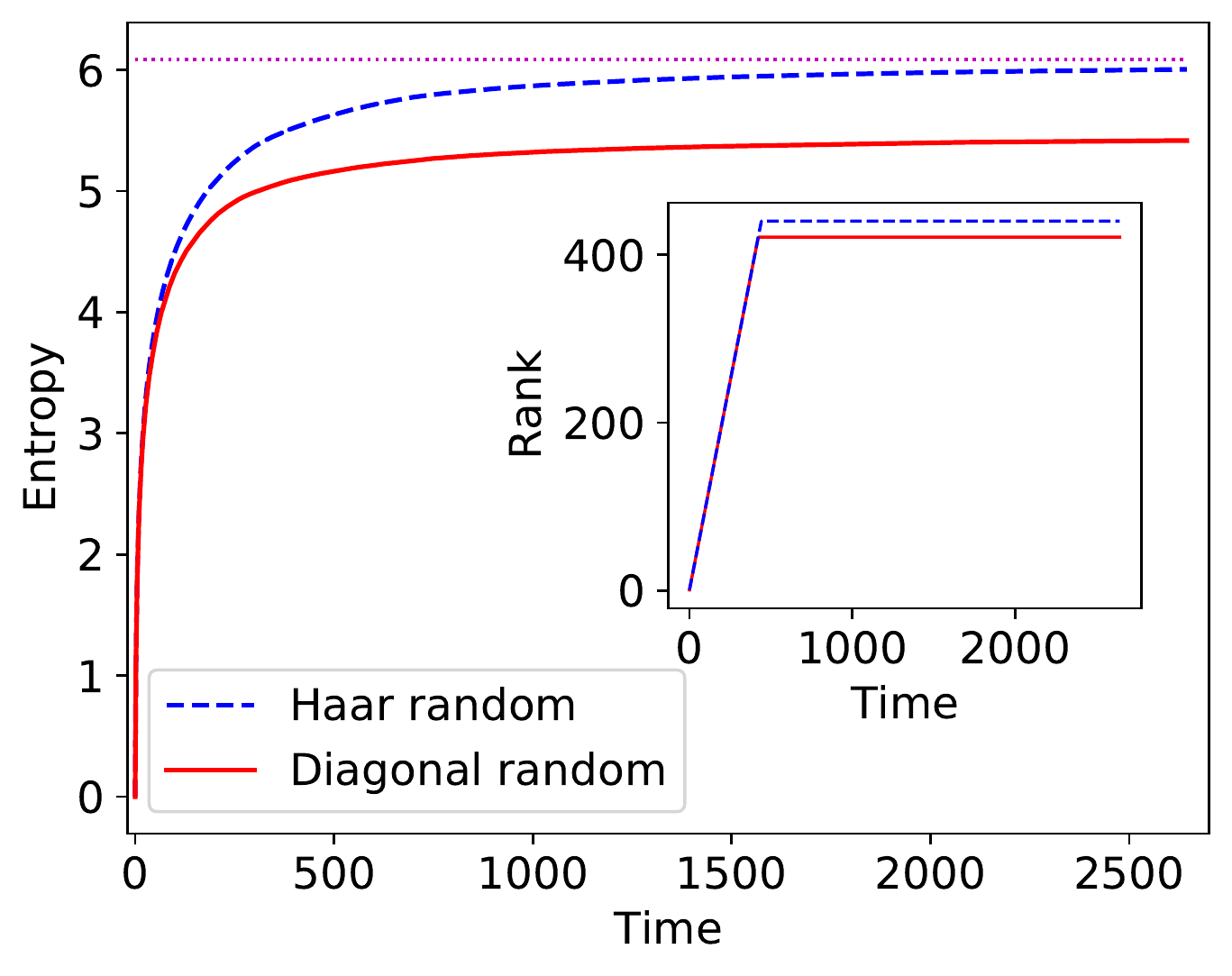}%
}
\caption{a) Comparison of collective  Fisher information of   random unitary  diagonal in a fixed basis with completely Haar random unitary (i.e., a different Haar random unitary at each time step) for $d=21$. The initial observable is $J_x$ and is evolved over time. The X-axis represents number of applications of the unitary map. The collective Fisher information, defined as inverse trace of the covariance metric ($1/\mathrm{Tr}{\bold{C}}$) quantifies the amount of information the measurement records have about the unknown parameters. b) Comparison of Shannon entropy of   random unitary,  diagonal in a fixed basis with completely Haar random unitary. As time passes,  operator space is more evenly sampled and Shannon entropy tends to saturate. The inset shows the rank of the covariance matrix. It is the dimension of the operator space spanned. Operator subspace of dimension $d-2$ is left out in the diagonal case, which is reflected in the entropy. The dotted line parallel to X-axis is $\Log(d^2-1)$, the maximum attainable entropy.}
\label{entr}
\end{figure*}
 Fisher information is closely related to other information metrics -- the mutual information $\mathcal{I}[\mbf{r};\mbf{\tilde{M}}]$ and fidelity \cite{MADHOK2016}.
 
Maximum information gain is obtained when all the eigenvalues of $\bold{C^{-1}}$ are equal \cite{MADHOK2016}. To get such an equal information gain in all the directions, the  operator dynamics needs to be unbiased. This encourages the  quantification of   the ``bias'' or ``skewness" in sampling, and Shannon entropy is a familiar metric that can achieve this.  Let us normalize  the eigenvalues of  $\mbf{C}^{-1}$  so that they become a probability distribution. As mentioned already, they represent the signal-to-noise ratios in  each direction. We can now calculate the Shannon entropy of this distribution $\mathcal{H}=- \sum \lambda_i \mathrm{log} \lambda_i$, where $\left\lbrace\lambda_i\right\rbrace$ is the set of normalized eigenvalues. With longer time  evolution, the initial observable that we started with traverses a trajectory, visiting all of the operator space that the unitary dynamics can span. If the dynamics is  unbiased, this would even out the eigenvalues which in turn maximizes the Shannon entropy.  Such an even sampling of the operator space gives high fidelity reconstruction for random pure states. This asymptotic saturation of entropy is evident in Fig. \ref{entr}b. i.e., Random unitary dynamics maximize information gain.

 \section{Statistical bounds on information gain}

In this section, we study the maximum information gain that can be generated in our tomographic protocol.
We notice that the inverse covariance matrix, $\bold{C^{-1}}= \bold{\tilde{\mathcal{O}}}^T\bold{\tilde{\mathcal{O}}}/\sigma^2$ has the form similar to a matrix from the Wishart-Laguerre ensemble \cite{wishart1928generalised,livan2018introduction}, 
  obtained  from rectangular matrix of real elements. The necessary condition for the covariance matrix to have eigenvalues that behave statistically like that of a Wishart matrix
 is to have uncorrelated and identically distributed matrix elements in the constituent matrices $\bold{\tilde{\mathcal{O}}}^T$ and $\bold{\tilde{\mathcal{O}}}$. Marchenko-Pastur distribution describes the behaviour of eigenvalues of the  Wishart matrices of the form $W^TW$, where $W$ are large rectangular random matrices with independent and identically distributed entries. For a Wishart matrix constructed from $D \times N$ rectangular random matrix with  $D \leq N$, the Marchenko-Pastur density function denoted by $\rho(\lambda)$ is given by
\begin{align}
\rho(\lambda) &= \frac{N}{2 \pi  \lambda} \sqrt{(\lambda-\lambda_-) (\lambda_{+}-\lambda)} \\
\lambda_{\pm}&= \frac{1}{N}\left(1- \left(\frac{D}{N}\right)^{-1/2}\right)^2
\end{align}
where $\lambda \in [\lambda_{-}, \lambda_{+}]$.  Note that in our protocol,  $\bold{C^{-1}}$ matrix is obtained by $\tilde{\mathcal{O}}^T\tilde{\mathcal{O}}/\sigma^2$, where
\begin{equation}
\tilde{\mathcal{O}}= \begin{pmatrix}
\tilde{\mathcal{O}}_{11} & \tilde{\mathcal{O}}_{12} &...&\tilde{\mathcal{O}}_{1d^2-1}\\
...&...&...&...\\
 \tilde{\mathcal{O}}_{N1} & \tilde{\mathcal{O}}_{N2} &...&\tilde{\mathcal{O}}_{Nd^2-1}
\end{pmatrix}
\end{equation}
Here $N$ is the total number of time steps and $d^2-1$ is the dimension of the operator space.
An element  $\tilde{\mathcal{O}}_{n\alpha}= \mathrm{Tr}(\mathcal{O}_n E_\alpha)$ is the expectation value of operator $\mathcal{O}_n$ along the direction $E_\alpha$ in the operator Hilbert space. Since the expectation value of measurements along each direction is obtained by averaging over a large number of identically prepared systems, $\tilde{\mathcal{O}}_{n\alpha}$ follows Gaussian distribution because of central limit theorem, with variance $\sigma^2$, the shot noise. Hence each element in $\tilde{\mathcal{O}}$ is identically distributed.

Now what remains is to prove that elements of  $\tilde{\mathcal{O}}$  are independent.  The successive operators $\lbrace\mathcal{O}_n\rbrace$  are obtained  by  conjugation action on the initial operator  by a Haar random unitary at each step. This makes the operators independent of each other  and hence the measurement values are uncorrelated upto one contraint, $N\lVert \mathcal{O}_0 \rVert^2=\sum_{i,\alpha} \tilde{\mathcal{O}}_{i\alpha}^2 ,$ where $\mathcal{O}_0$ is the initial operator.  However, when $N$ and $d^2-1$ are large, $\tilde{\mathcal{O}}$ behaves effectively as a random matrix with independent and identically distributed entries.  Now we numerically demonstrate that the Haar random evolution accomplishes this.
$\bold{C^{-1}}$ is a $d^2-1$ dimensional full rank matrix.  As the dimension of the Hilbert space tends to be very large, the eigenvalues  of Wishart matrices become continuous and follow the Marchenko-Pastur density function as seen  in Fig.  \ref{fig:loe}.

We estimate the collective Fisher information using the Wishart-Laguerre ensemble.  Let $\left\lbrace\lambda_i\right\rbrace_{i=1}^D $ be the eigenvalues of $\bold{C^{-1}},$ where $D=d^2-1$.  In the limit of large $N$, we can approximate the sum by an integral.
\begin{equation}
FI = \frac{1}{\sum_i^D \frac{1}{\lambda _i}} \approx \frac{1}{D \int \frac{1}{\lambda} \rho(\lambda)d \lambda} = \frac{1}{D \langle \frac{1}{\lambda}  \rangle} \label{fi}
\end{equation}   
where $\rho(\lambda)$ is the Marchenko-Pastur density. In our numerical simulation, we evolved the system for $6d^2$ time steps and used the measurement records obtained to generate the Wishart matrix. Therefore in our simulations, the  parameters  in the density function are $D=440$ and $N=2646.$  
The collective Fisher information obtained using the integral approximation is 18.03907, in excellent agreement with the Haar random case. Using the covariance matrix of Haar random evolution,  we get $FI$=18.76207, after 2646 time steps.  The small difference in the values obtained can be attributed to the dimension being small for the eigenvalues to be continuous.


To quantify the bias in the operator space dynamics, we calculate the Shannon entropy. We normalize the eigenvalues of $\bold{C^{-1}}$  so that they form a probability distribution  and  compute the the Shannon entropy $\mathcal{H}=-\sum \ \lambda_i \Log\lambda_i$, where $\lbrace \lambda_i \rbrace$ are the normalized eigenvalues. For the  Haar random case, the { average} entropy numerically obtained for an ensemble of random states of dimension $d=21$, after 2646 iterations, which we denote by subscript ``$rs$",    is $\mathcal{H}_ {rs}=6.00415$. The Shannon entropy of Wishart-Laguerre orthogonal ensemble, denoted by subscript ``$loe$" can be calculated as  
  \begin{equation}
  \mathcal{H}_{loe}= - D\int_{\lambda_-}^{\lambda_+}\lambda \Log \lambda  \frac{N}{2 \pi  \lambda} \sqrt{(\lambda-\lambda_-) (\lambda_{+}-\lambda)}d\lambda\label{mpdensity}
\end{equation}   
Using this integral approximation, which works better for large dimensions, we get $\mathcal{H}_{loe}=6.00363$ in remarkable agreement with the Haar random case. Fig. \ref{fig:loe} shows the distribution of normalized eigenvalues of  $\bold{C^{-1}}$ for Haar random evolution and the Marchenko-Pastur density function. When all the  eigenvalues are equal,  the expected Shannon entropy  for  $d=21$, is $\mathcal{H}_{exp}= \log{(d^2-1)}= 6.08677$. Indeed, an application of a different random unitary at each time step is the most unbiased dynamics we can hope to perform.

\begin{figure}[hbtp]
	\centering
	\includegraphics[width=8cm,height=5.5cm,angle=0]{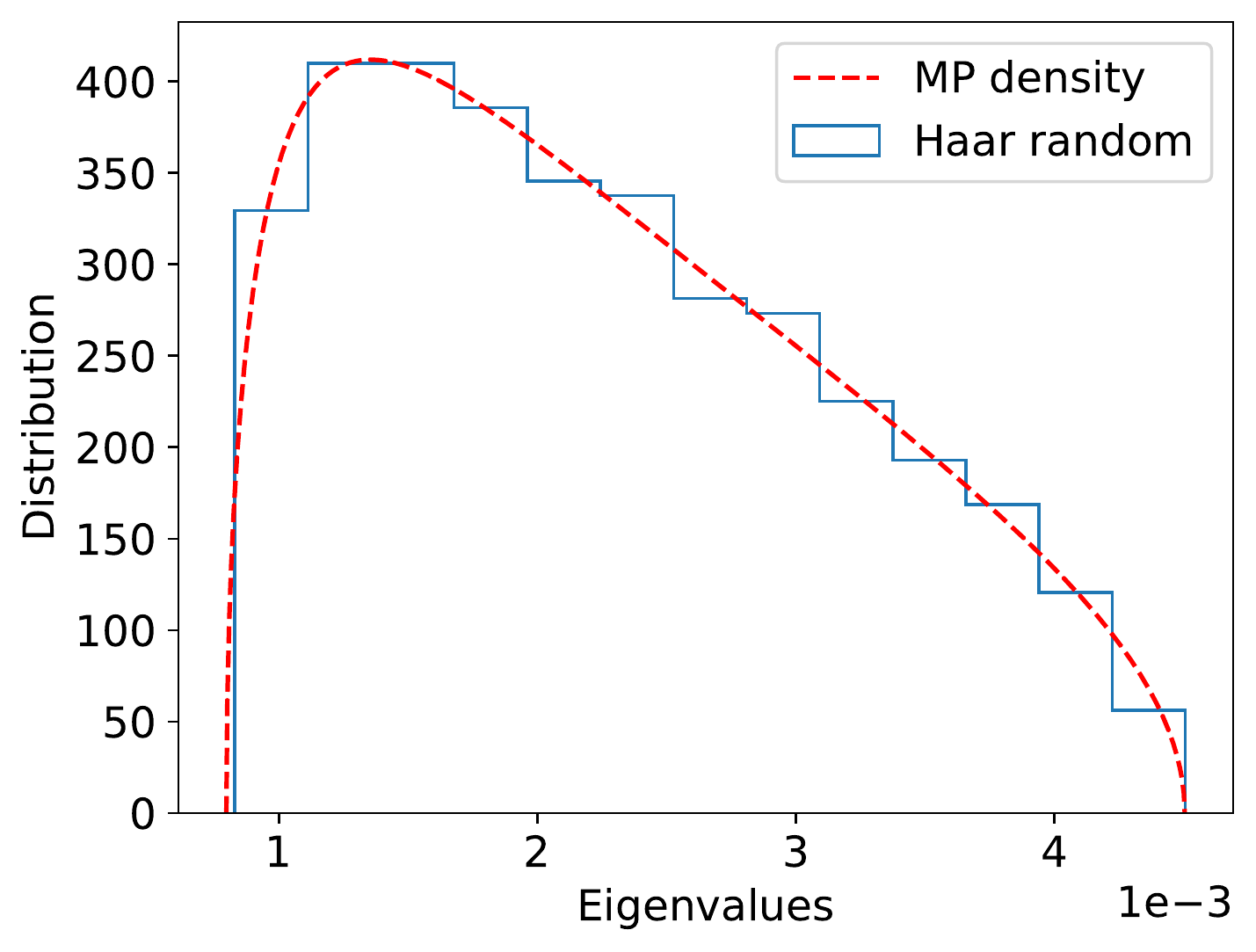}
	\caption{Histogram of eigenvalue distribution of $\bold{C^{-1}}$ of Haar random evolution and Marchenko-Pastur density function. Remarkable agreement is seen in the eigenvalue distribution.} \label{fig:loe} 
\end{figure}

\subsection*{Information gain for random diagonal unitaries}

In this section, we study the maximum information gain that can be generated in our tomographic protocol through the application of random diagonal operators. For this case, the  inverse covariance matrix does not obey the Marchenko-Pastur distribution.
 In the standard Hilbert space, the operator we apply at any time step is of the form, $U = \sum_{j=1}^d  e^{-i\phi_j}\ket{j}\bra{j}$,  where $e^{-i\phi_j}$ and $\ket{j}$ are its eigenvalues and eigenvectors, respectively. Since we will be indexing the unitaries as before,
we can rewrite this as $U_{m}=\Sigma_{j=1}^d e^{-i \phi_{mj}} \ket{j} \bra{j} $, where $U_{m}$ is the random diagonal unitary applied at time step $m$.
Here the exponential phase factors $\phi_{mj}, \in [0, 2\pi]$, is chosen uniformly at random.  
Therefore, after $n$ time steps, 
\begin{equation}
\mathcal{O}_n = U_{n}^{\dagger}(\Delta t)U_{n-1}^{\dagger}...U_{1}^{\dagger}(\Delta t)\mathcal{O}_0U_{1}(\Delta t)U_{2}(\Delta t)...U_{n}(\Delta t) \label{0003}
\end{equation}
 Now we use the superoperator picture~\cite{caves1999quantum, madhok2014information}. The superoperator of the measured observable after $n$ times steps, $|\mathcal{O}_n)= \bold{U_n U_{n-1} ... U_{1}} |\mathcal{O}_0)$, where the superoperator map is $\bold{U} = U^\dagger \otimes U^{T}$. Using this,
we can write our unitary superoperator map explicitly as $\bold{U} = \sum_{j, k}^d  e^{-i(\phi_k - \phi_j)}|j,k)(j,k|$, where one defines $|j,k) = \ket{j}\otimes\ket{k}^*$, with $^*$ denoting complex conjugation. Therefore in our notation, $ \bold{U_n U_{n-1} ... U_{1}} =  \sum_{j, k}^d  e^{\sum_{m=1}^n {-i(\phi_{mk} - \phi_{mj})}}|j,k)(j,k|$.

  In the superoperator representation, after $N$ time steps, the inverse of this covariance matrix is  $\bold{C^{-1}} = \sum_{n=1}^N |\mathcal{O}_n) (\mathcal{O}_n|$. We can write this as
\begin{equation}
\label{eq:InvCExp}
\bold{C^{-1}} = \sum_{j,k=1}^d\sum_{j',k'=1}^df_{j,k}^{j',k'}(j',k'|\mathcal{O}_0)(\mathcal{O}_0|j,k)|j',k')(j,k|,
\end{equation}
where,
 \begin{equation}
f_{j,k}^{j',k'}=\sum_{n=1}^N e^{\sum_{m=1}^n {-i(\phi_{mj} - \phi_{mj'}-\phi_{mk}+\phi_{mk'})}} 
\end{equation}
which can be simplified as in (\ref{on})
 \begin{equation}
f_{j,k}^{j',k'}=\sum_{n=1}^N e^{ {-i(\Phi_{nj} - \Phi_{nj'}-\Phi_{nk}+\Phi_{nk'})}} 
\end{equation}

{Notice that if $\Phi_{nj}-\Phi_{nj'} - \Phi_{nk}+\Phi_{nk'}=0$, $\forall n$, for a particular choice of $j,k,j',k'$, the quantity $f_{j,k}^{j',k'}=N$. For an arbitrary unitary map $U$, we will assume that the only way this can happen is if $(j=k) \wedge (j'=k')$ or $(j=j')\wedge (k=k')$, which is certainly true for a random unitary. }

With this assumption, we can approximate $\bold{C^{-1}}$ by terms that scale with $N$ in the large $N$ limit. The inverse covariance matrix is

\begin{widetext}
\begin{equation}
\label{inverseC}
\bold{C^{-1}} \approx N\left[\sum_{j,k=1}^d|(j,k|\mathcal{O}_0)|^2|j,k)(j,k|+ \sum_{j\ne k=1}^d (j,j|\mathcal{O}_0)(\mathcal{O}_0|k,k)|j,j)(k,k| \right] .
\end{equation}
\end{widetext}
{In this superoperator representation, $\bold{C^{-1}}$ is a $d^2 \times d^2$ dimensional matrix with $d^4$ elements in total. Notice that on the right-hand side of (\ref{inverseC}), the total number of terms are only of order $d^2$.  Therefore the matrix $\bold{C^{-1}}$ is sparse for large $d$.  The degree of sparsity increases with the increase in the dimension of the space. The first sum in (\ref{inverseC}) contains the diagonal elements.  In the limit of large $d$, eigenvalues of $\bold{C^{-1}}$ are very close to the diagonal terms, because of the limited interaction with other elements in the matrix.  Since the inverse covariance matrix is a superoperator in a real vector space of $d^2$ dimensions, let us compare the normalized eigenvalues with the Porter-Thomas distribution \cite{wootters1990random}. The motivation for this comparison is that $\bold{C^{-1}}$ can be thought of as being picked from a unitarily invariant measure by its construction, with real eigenvalues. }}
\begin{figure}
\centering
\includegraphics[width=8cm,height=5.5cm,angle=0]{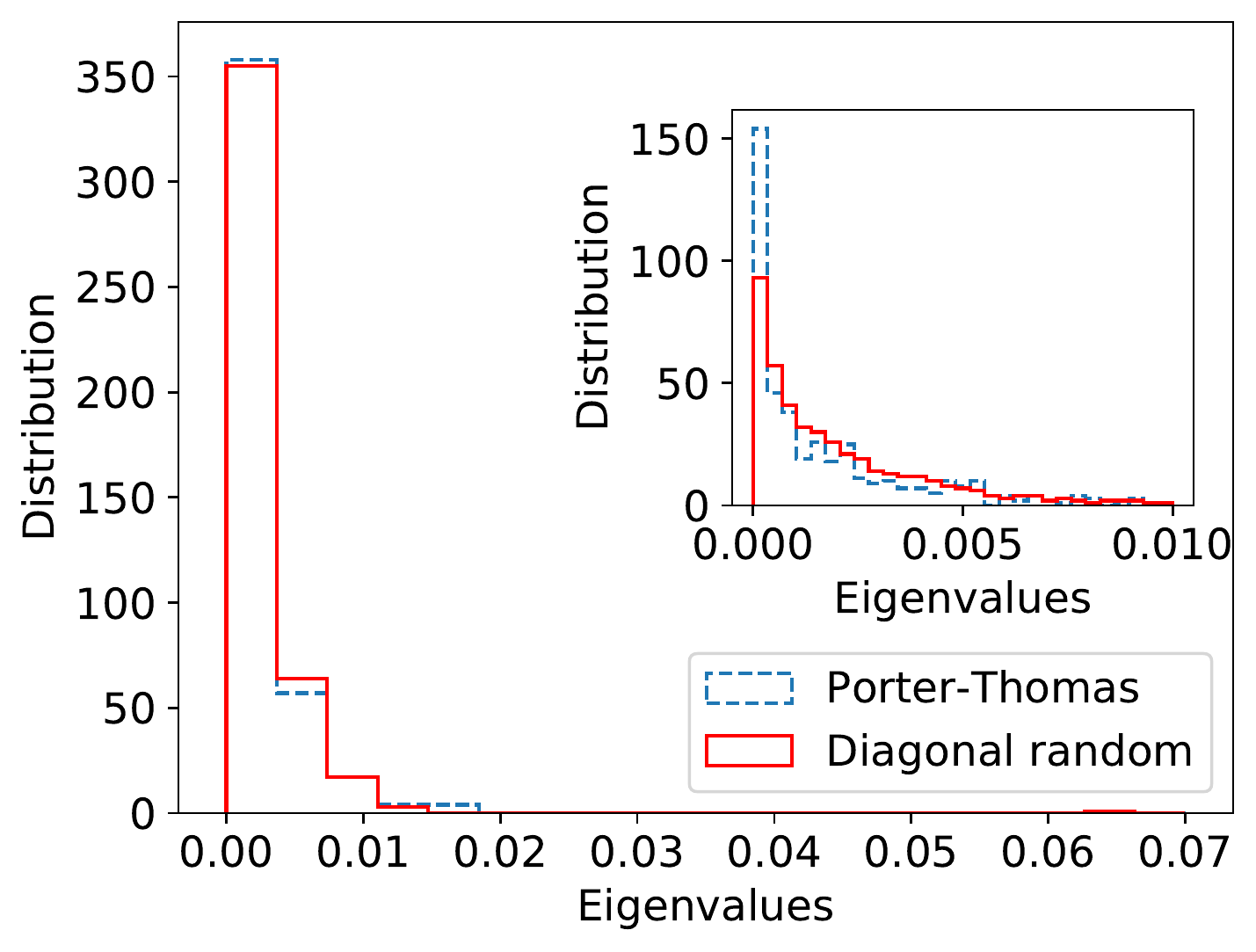} 
\caption{Distribution of eigenvalues of $\bold{C^{-1}}$ for the diagonal random case, along with the distribution of   random numbers generated from  Porter-Thomas distribution. The inset shows the zoomed in distribution for very small eigenvalues.}
\label{ptfig}
\end{figure}
{The Porter-Thomas distribution given below in (\ref{pt}) represents frequency distribution of components of a pure unit vector, chosen uniformly at random in a $d^2$- dimensional real Hilbert space. Let $a_i$ be the $i^{th}$ component of the random real pure state, then  probability for obtaining the $i^{th}$ outcome $p_i= a_i^2$. When the dimension of the Hilbert space $d^2$ is large,  the  $i^{th}$ outcome occurs  $\lambda_i= d^2p_i$ times. The distribution of these frequencies follow the Porter-Thomas distribution
\begin{equation}
\rho(\lambda)= \frac{1}{\sqrt{2\pi \lambda}} e^{-\lambda/2} \label{pt}
\end{equation}
  We denote the Shannon entropy obtained from Porter-Thomas distribution by subscript $``pt"$.
\begin{equation}
\mathcal{H}_{pt}=-d^2 \int_0^\infty \frac{\lambda}{d^2} \Log \left(\frac{\lambda}{d^2}\right) \frac{1}{\sqrt{2\pi \lambda}} e^{-\lambda/2} dx=5.35941\label{enpt}
\end{equation}
Also using properties of random states in a real vector space \cite{wootters1990random}, the expected entropy of pure states with real coefficients,  $\mathcal{H}_{exp}=\Log(d^2) -0.729637 $, gives $5.35941$. Both these values obtained for entropy are very similar, and  in very good agreement with $\mathcal{H}_{rs} = 5.41684$ as obtained by our numerical simulations using random diagonal unitaries.  Further evidence of this is given in Fig. \ref{ptfig}  which compares the eigenvalues of $\bold{C^{-1}}$ with the Porter-Thomas distribution. In very high dimensions, sparsity of the matrix is so high that  all the correlations die, and eigenvalues of $\bold{C^{-1}}$ form a truly random vector. In that asymptotic limit, eigenvalues  follow the Porter-Thomas distribution.
}

{Now  let us look at the rate of information generation.  The Fisher information obtained for the diagonal random case in our numerical simulation is $5.37541$ after $N=2646$ time steps. Using the constraint $\Tr(\bold{C^{-1}})= N ||\mathcal{O}_0||^2$ to re-scale the numbers generated according to Porter-Thomas distribution, so that they are in same footing with the numerical case,
\begin{equation}
\lambda_i \rightarrow \frac{\lambda_i}{d^2} \Tr(\bold{C^{-1}})+ d^2
\end{equation} 
where $d^2$, has been added as a constant regularization factor which we also had in our simulations to avoid infinities while finding the Fisher information. 
\begin{equation}
FI= \frac{1}{\left[d^2 \int_0^{\infty} \left(\frac{d^2}{\lambda \Tr(\bold{C^{-1}})+ d^4 } \right)\frac{1}{\sqrt{2 \pi \lambda}}e^{-\lambda/2} d\lambda\right] }  \label{fipt}
\end{equation}
which yields $4.31174$, a low value compared to the one we obtained numerically. This is because Porter-Thomas distribution is heavily populated by very small numbers.  $\bold{C^{-1}}$ has a lesser number of very small eigenvalues as seen in the inset of Fig. \ref{ptfig}. Therefore the eigenvalues of $\bold{C}$, which is the set of inverted eigenvalues $\left\lbrace 1/\lambda_i\right\rbrace$ has more smaller numbers than the  corresponding Porter-Thomas set.  Hence  the trace of the covariance matrix is smaller and  the $FI$ larger, compared to the Porter-Thomas case.} 

 \section{Continuous measurement tomography and it's connection to quantum chaos and spectral statistics}

Our results of the previous section indicate a connection between random diagonal unitaries and quantum chaos. One can characterizes quantum chaos dynamically, by ``ergodic mixing", i.e., something that takes a localized state in phase space and maps it to a random state, smeared across phase space.  As shown in previous literature, a quantum chaotic map takes a localized state to a pseudorandom state in Hilbert space.  This is characterized by the entropy production of the probability distribution with respect to the standard basis \cite{BandyopadhyayArul2002, Lakshminarayan, Bandyopadhyay04, Zyczkowski1990, ScottCaves2003, tmd08}. Intuitively,  information gain in quantum tomography is a closely related phenomenon where ergodic mixing due to chaos can be viewed in the Hiesenbeg picture and interpreted
as the rate of obtaining information in different directions of the operator space.

In contrast, a common approach is to characterize chaos using static properties.  In this approach, the signature of chaos in a quantum system is in the energy level statistics of the Hamiltonian (or phases of a Floquet map).  Depending on the symmetries, quantum chaotic systems are classified as Gaussian or Circular Orthogonal (GOE/COE), Gaussian or Circular unitary GUE/CUE and Gaussian or Circular Symplectic Ensembles (GSE/CSE) \cite{Haake}. 

The question then is, does ergodic mixing depend sensitively on the eigenvalues of the Hamiltonian $H$ or Floquet map $\mathcal{U},$ or just on the eigenvectors?  
Is the power of $\mathcal{U}$ to generate randomness related to its eigenvectors, and not eigenvalues?   Are any two Hamiltonians or Floquet maps with the same eigenvalue spectrum the same?    The physical Hamiltonians for regular systems will have nonrandom eigenvectors as well as level statistics corresponding to a Poissonian distribution. 

To decouple the role played by eigenvalues and eigenvectors of the dynamics in the rate of information gain in tomography, we construct quantum maps that have an eigenspectrum corresponding to regular systems and eigenvectors that are random with respect to a standard basis. This is obtained by doing a unitary transformation to a given eigenbasis. The other possibility of regular eigenvectors but eigenspectrum exhibiting level repulsion- a signature of chaos, is also considered.

\begin{figure}[hb]
	\centering	\includegraphics[width=8cm,height=5.5cm,angle=0]{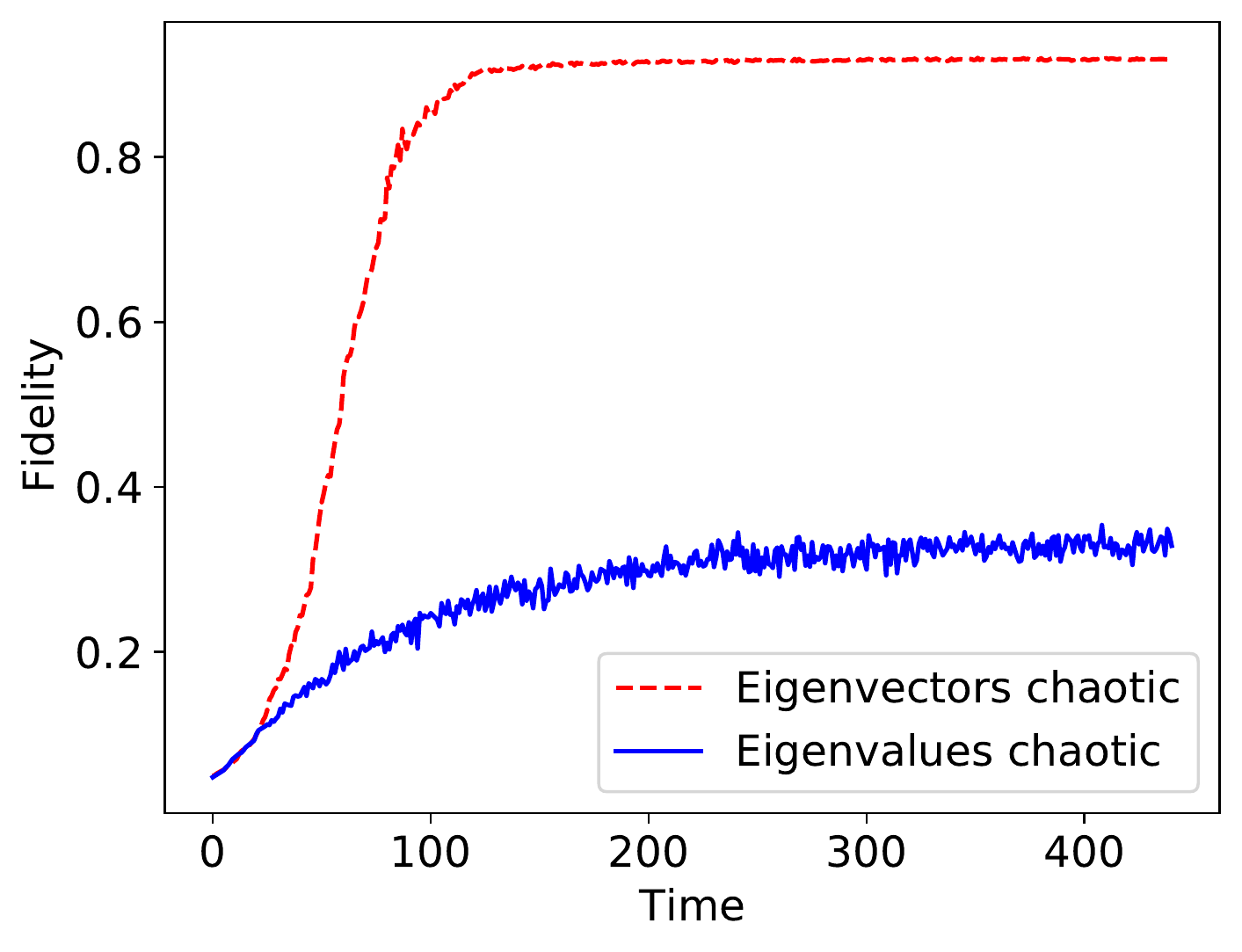}
	\caption{Average reconstruction fidelity over ten random states of $d=21$ using repeated application of a kicked top unitary, $\mathcal{U}= e^{-i1.4 J_x} e^{\frac{-ik_0}{(n-1)} J_z^2}$.  $k_0$ is the chaoticity parameter of the map. The $X$ axis shows the number of applications of the Floquet map. The `eigenvalues chaotic' case is when eigenvalues are picked from the floquet in the chaotic regime with chaoticity 7 and eigenvectors are picked from the floquet with chaoticity 0.5. The other case follows similarly. } \label{fig:fisher} 
\end{figure}

\begin{figure*}[!htb]
\centering
	\subfloat[\label{sfig:testa}]{%
		\includegraphics[width=8cm,height=5.5cm,angle=0]{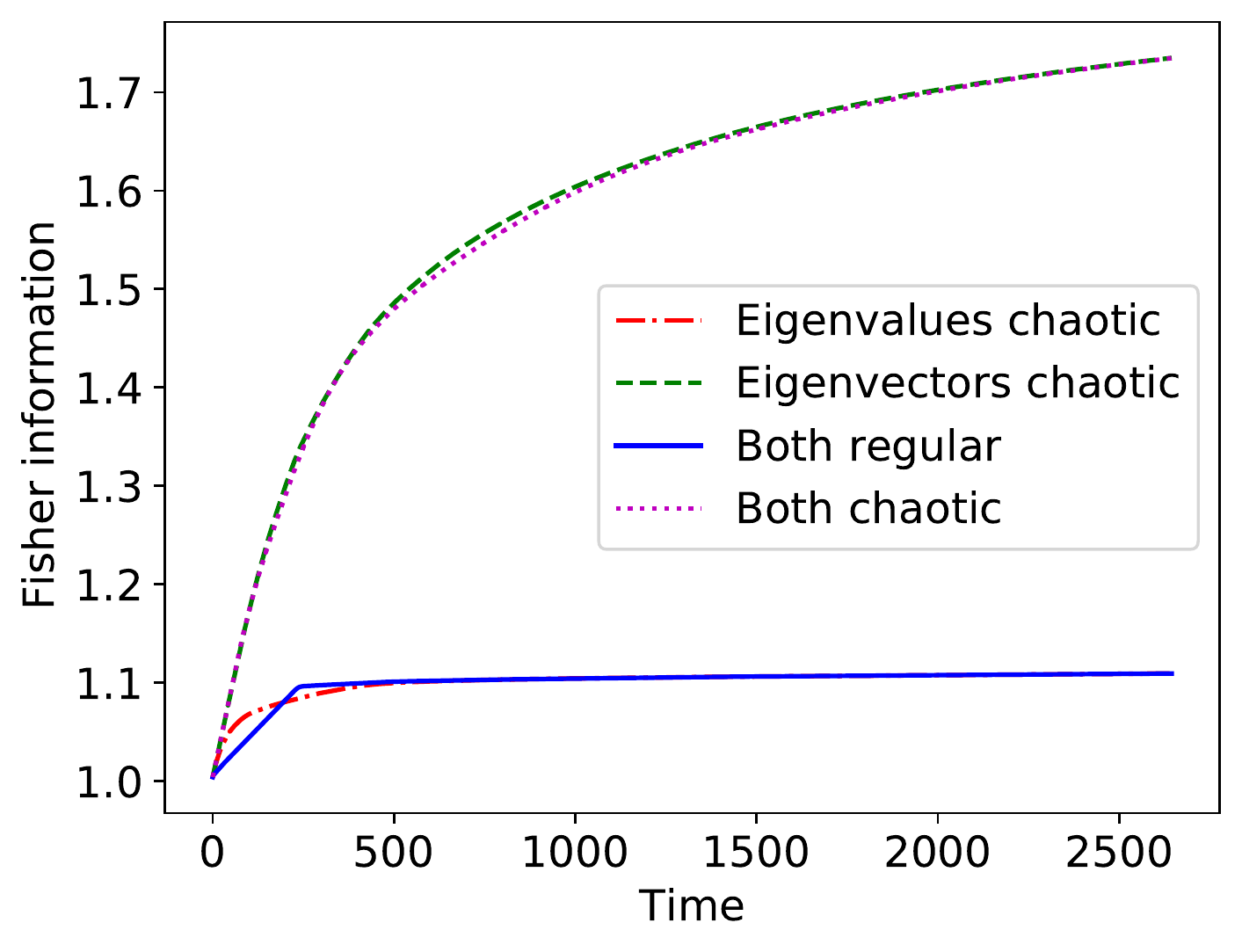}%
	}\hfill
	\subfloat[\label{sfig:testa}]{%
		\includegraphics[width=8cm,height=5.5cm,angle=0]{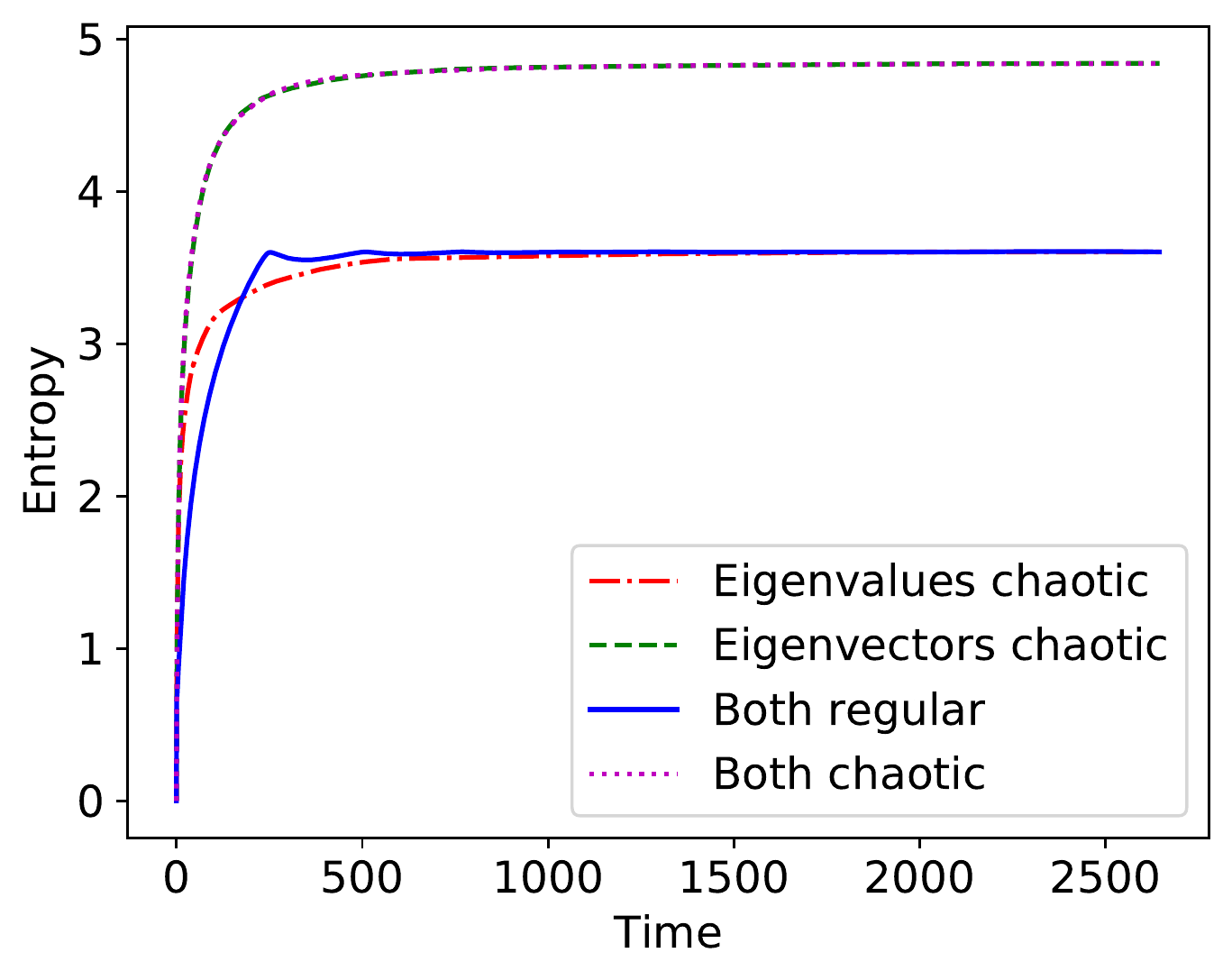}%
	}
	\caption{{The initial observable is $J_z$ and is evolved over time using repeated application of kicked top flouquet operator, $\mathcal{U}= e^{-i1.4 J_x} e^{\frac{-ik_0}{(n-1)} J_z^2}$ for $J=10$.  $k_0$ is the chaoticity parameter. The $X$ axis shows the number of applications of the Floquet map.  The `eigenvalues chaotic' case is when eigenvalues are picked from the floquet in the chaotic regime with chaoticity 7 and eigenvectors are picked from the floquet with chaoticity 0.5. Other cases follow similarly. We observe that when eigenvectors are spread out with more support in the Hilbert space, information gain is  more.  b) Comparison of Shannon entropy with  kicked top evolution for various cases described in part a.  The difference in the saturation value means that when eigenvectors are picked from a  chaotic unitary, they span more operator space.}}
	\label{floquet}
\end{figure*}

\begin{figure*}[!htb]
	\subfloat[\label{sfig:testa}]{%
		\includegraphics[width=8cm,height=5.5cm,angle=0]{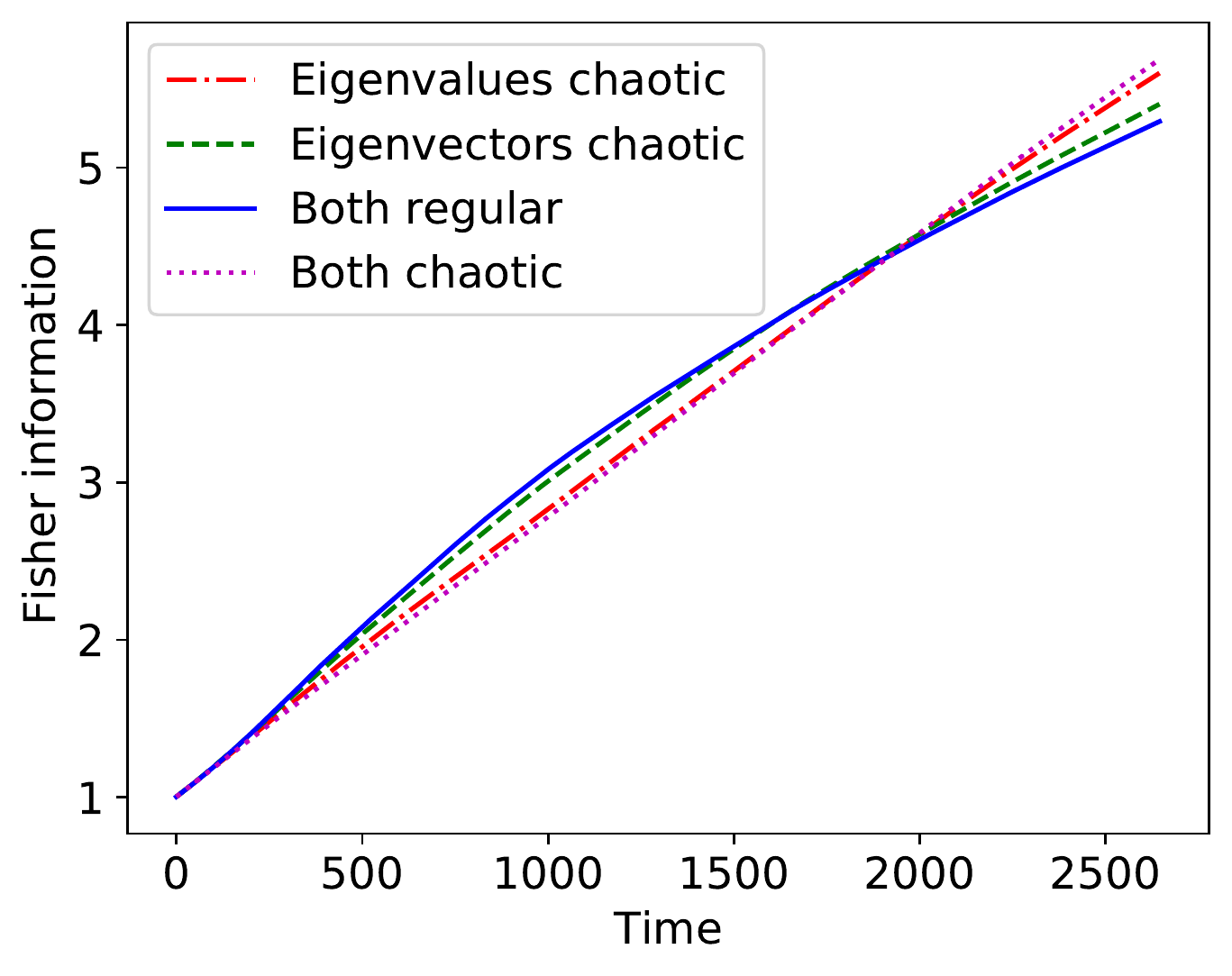}%
	}\hfill
	\subfloat[\label{sfig:testa}]{%
		\includegraphics[width=8cm,height=5.5cm,angle=0]{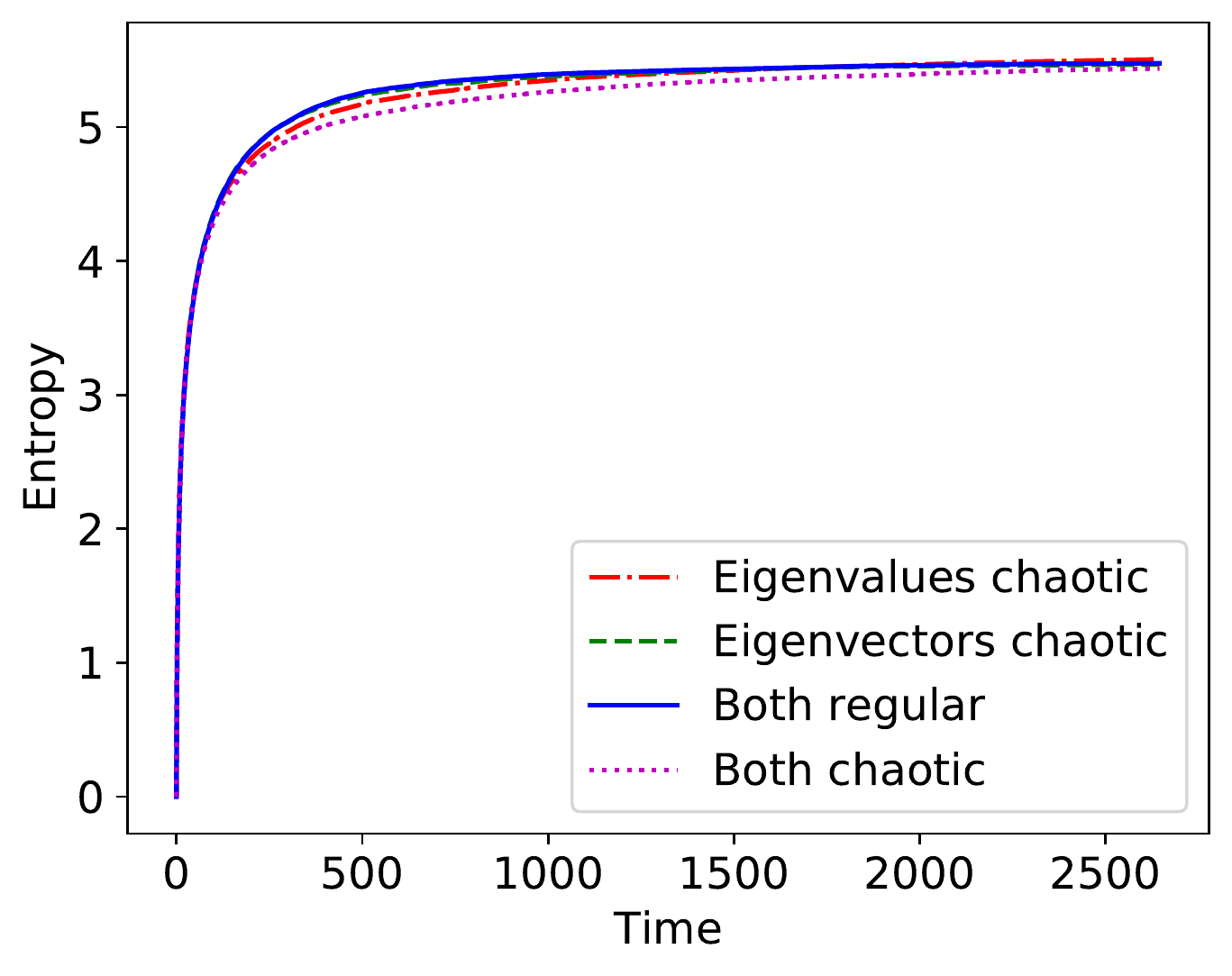}%
	}
	\caption{{ When the initial operator $J_z$ is rotated by a random unitary to $U^{\dagger} J_z U$, the differences seen in the information generation and entropy in the previous figure disappear. The $X$ axis shows the number of applications of  the Floquet map. A larger saturation value of Shannon entropy shows that more of the operator space is spanned, much more than the previous case. }}
	\label{floquet1}
\end{figure*}

 {To this end, we use the repeated application of the kicked floquet to evolve the initial operator $J_z.$  Results are shown in Fig. \ref{fig:fisher} and Fig. \ref{floquet}. There is a stark increase in the achieved fidelity when eigenvectors are chosen from a chaotic kicked top unitary.  Figure \ref{floquet} shows that when eigenvalues of the unitary are non-degenerate, the rate of information generation and the amount of operator space spanned during the evolution are solely dependent on the nature of eigenvectors.   Choosing the eigenphases from a chaotic unitary doesn't give any advantage in this case. Figure \ref{floquet1} shows the evolution of the same system with a rotated initial operator.}

From these observations, it is clear that the eigenvalue statistics of  $\mathcal{U}$, which is a basis-independent criterion of quantum chaos, is not necessary for the information gain in tomography. This suggests that it is the RMT statistics of the eigenvectors of $\mathcal{U}$ that is responsible for faster quantum state reconstruction.
Randomizing the eigenvectors of the initial operator has resulted in washing away the differences seen in the rate of information generation and entropy, in Fig. \ref{floquet}.  This again demonstrates the basis dependence  of randomness generation.
The message our study imparts is that in general,  the dynamical signatures of chaos, like the generation of near maximally entangled random states and information gain in tomography, are a basis-dependent feature of a system. Either we need initially random operators that might be hard to implement, or one needs a dynamics
that gives rise to pseudorandomness in operator space that generates observables that has support over 
almost the entire $d^2-1$ dimensions.

 Figure \ref{entr} shows a faster growth of Fisher Information 
 as compared to the case to Fig. \ref{floquet} for the kicked top. As discussed above,
 different Haar random unitaries saturate the full $d^2 -1$ dimensional operator space, whereas 
a repeated application of a single Haar random unitary (like the kicked top) misses $d-2$ dimensions. This manifests in the values of Fisher Information.
The information gain when one employs different Haar random unitaries is naturally more rapid as compared to being restricted to repeated application of a single kicked top which is further restricted by additional constraints that we describe below.

In addition, the kicked top has a parity symmetry {given by $R = \exp(-i\pi j_x)$}. In the basis {in which} the parity operator {is diagonal}, the Floquet map has a block diagonal structure {corresponding to the $+1$ and $-1$} parity eigenvalues.
The parity operator $R = e^{-i \pi J_x}$ commutes with the kicked top unitary, $\mathcal{U}$(supplementary section in \cite{madhok2014information}). Therefore, there exists a basis in which both $J_x$ and $\mathcal{U}$ are diagonal. For parameters in which the classical dynamics is globally chaotic, we{, in general,} expect the Floquet operator to have the statistical properties of a random matrix chosen from the circular orthogonal ensemble (COE)~\cite{Haake}.
{Because of the additional parity symmetry, we must choose a block diagonal matrix whose blocks are sampled from the COE in the basis in which the parity operator $R$ is diagonal, thus having the same block structure as the Floquet map}.  This is the reason that the saturation value of Shannon entropy in Fig. \ref{floquet} ($4.8398$) is lower than that of the random diagonal case Fig. \ref{entr} (where Shannon entropy reached 5.41684).  That means the Floquet dynamics spans a smaller subspace of the operator space.  

\section{Discussion}

Quantum tomography is a resource-intensive process of fundamental importance in quantum information theory. The challenge is to accomplish this in an efficient manner and research is focused on optimizing  protocols. Many techniques like compressed sensing \cite{gross2010quantum} or the taking advantage of the positivity constraint \cite{kalev2015power} focus on the prior information available in state estimation. In this work, we took a different approach and showed that with a dynamics that is not informationally complete, we can still get very high fidelities in quantum state reconstruction.
In particular, what we have seen in this paper is that random unitaries diagonal in a particular basis do almost as good as Haar random unitaries in terms of fidelity of  state reconstruction. This is despite missing out on the information from $d-1$ dimensional subspace of the operator space. We quantified the rate of information gain using collective Fisher information and used Shannon entropy to quantify uniformity in operator sampling. We gave statistical bounds on information gain and also discussed how close diagonal random unitary dynamics come in saturating these bounds.
Finally, we saw that asymptotic evolution using  Haar random unitaries is modeled remarkably well by the Wishart-Laguerre orthogonal ensemble. We also obtained an intuitive understanding of the vector space visited by the random unitary maps considered. Thus our work is an important contribution towards the applications of random matrix theory in quantum information.

One interesting question that arises from our study is the performance of quantum process tomography using states generated by random diagonal unitaries as inputs. Quantum process tomography is the process of determining the trace-preserving completely positive map that is applied on the system and therefore $d^4-d^2$ real numbers are required to completely characterize it.
Using continuous measurement quantum process tomography, how close does a random dynamics or random dynamics diagonal in a fixed basis come in getting an accurate description of the map?

The flip side of quantum tomography is quantum control. One requires an informationally complete set for perfect state reconstruction. Similarly, such an informationally complete dynamics
will be able to steer an initial state to any target state in the Hilbert space.  
 The ability of random diagonal unitaries to generate information in $d^2-d+1$ dimensions tells us as to what target states are achievable starting from a fiducial state.  
 
 Randomized
benchmarking is used to 
estimate the fidelity between the applied map
and the target unitary, in the presence of errors. How well do randomized benchmarking protocols work when one only has random diagonal unitaries at disposal? Is there a way to perform randomized benchmarking with a restricted set of unitaries? These are questions we like to address in the future.

\section{Acknowledgement}
We are grateful to Arul Lakshminarayan, Ivan Deutsch and Prabha Mandayam for useful discussions.


\clearpage
\bibliographystyle{unsrt}
\bibliography{references1}

\end{document}